\documentclass{jfm}
\usepackage{graphicx}
\usepackage{epstopdf, epsfig}
\usepackage{graphicx}
\usepackage{epstopdf,epsfig}
\usepackage{newtxtext}
\usepackage{newtxmath}
\usepackage{natbib}
\usepackage{newtxtext}
\usepackage{newtxmath}
\usepackage{amsmath}
\usepackage{multirow}
\usepackage{tabularx}
\usepackage{physics}
\graphicspath{./plots/}
\usepackage{caption}
\usepackage{subcaption}

\shorttitle{Polydisperse Particle-laden Flow Instability}
\shortauthor{Z. Liu, Y. Dagan}

\title{Generalized stability theory of polydisperse particle-laden flows. Part1. Channel flow}

\author{Zhixuan Liu\aff{1}
 \and Yuval Dagan\aff{1} \corresp{\email{yuvalda@technion.ac.il}}}

\affiliation{\aff{1}Faculty of Aerospace Engineering, Technion - Israel Institute of Technology, Haifa, 3200003, Israel}

\begin{document}

\maketitle

\begin{abstract}
We present a generalized hydrodynamic stability theory for interacting particles in polydisperse particle-laden flows.
The addition of dispersed particulate matter to a clean flow can either stabilize or destabilize the flow, depending on the particles' relaxation time-scale relative to the carrier flow time scales and the particle loading. 
Here, to study the effects of polydispersity and particle interactions on the hydrodynamic stability of shear flows, we propose a new generalized mathematical framework by combining a linear stability analysis and a discrete Eulerian sectional formulation to describe the flow and the dispersed particulate matter. 
In this formulation, multiple momentum and transport equations are written for each size-section of the dispersed phase, where interphase and inter-particle mass and momentum transfer are modelled as source terms in the governing equations.
A new modal linear stability framework is derived by linearizing the coupled equations, leading to a generalized eigenvalue problem incorporating the carrier flow equations and terms for each size section of the dispersed phase.
Using this approach, particle-flow interactions, such as polydispersity, droplet vaporization, condensation, and other interphase mass transfer interactions, may be modelled. 
The method is validated with linear stability analyses of clean and monodisperse particle-laden flows. 
We show that the stability characteristics of a channel flow laden with particles drastically change due to polydispersity.
We find that while relatively large monodisperse particles of moderate Stokes numbers tend to stabilize the flow, adding a very small mass fraction of low-to-moderate Stokes number particles may significantly increase the perturbation growth rates, and for high-Reynolds numbers may destabilize flows that might have been regarded as linearly stable in the monodisperse case. Equivalent Stokes numbers for such bi-sectional particle distribution are devised, which enables the first comparison between monodisperse and polydisperse flow instability.
These findings may apply to a vast number of fluid mechanics applications involving particle-laden flows such as atmospheric flows, particle imaging velocimetry, dusty flows, environmental flows, medical applications, propulsion, and energy systems.
\end{abstract}


\section{Introduction}
\label{sec:intro}
Interactions between unsteady flows and dispersed particulate matter may lead to instability. The problem of turbulent particle-laden flow involves multi-material, multiphase, and multiple scales. Particles may either stabilize or destabilize a flow based on the properties of the particles and the carrier fluid~\citep{saffman1962stability}. Particle interactions may alter the transition from laminar to turbulent flow and modulate turbulence of fully developed flows. However, the transition and turbulence modulation mechanisms by interacting particles remain elusive. Due to complex inter-phase and particle interactions, the dispersed phase significantly complicates the mathematical formulation and computation of multiphase flows. Despite the fact that computational methods have improved considerably, most existing computational models make use of simple analytical laws and empirical correlations and heavily rely on a significant separation of scales. Moreover, the impact of particle interactions on the hydrodynamic stability is rarely accounted for, neither in high-fidelity simulations nor in theoretical models.

Linear stability analysis of particle-laden shear flows was first conducted by \cite{saffman1962stability}, who proposed that the addition of solid particles can either stabilize or destabilize the carrier flow, depending on the particles' relaxation time-scale relative to the carrier flow time scales.
He showed that when the particle relaxation time is short compared to the characteristic time scale associated with the flow, then the addition of particles may destabilize a gas flow. In contrast, if the particle relaxation time is relatively large, then the particles will have a stabilizing effect.
Since then, even though the transition from laminar to turbulent flow has been studied extensively~\citep{drazin1982hydrodynamic}, the influence of particulate matter in dispersed multiphase flows on the transition to and modulation of turbulence is still not fully understood~\citep{balachandar2010turbulent, matas2003transition, klinkenberg2011modal}. 

Recently,~\cite{wang2019modulation} showed, using DNS of inertial particles suspended in turbulent flow, that high-inertia particles greatly weaken the large-scale vortices, whereas low-inertia particles strengthen the large-scale vortices and reduce the critical Reynolds number.
\citet{klinkenberg2011modal} investigated the effect of particles on the modal and non-modal stability of channel flows. In their study, the critical Critical Reynolds numbers in particle-laden channel flows with different particle loadings depend on the particle's dimensionless relaxation time. Particles of intermediate dimensionless relaxation time $S$ result in the highest critical Reynolds numbers, while particles on the two extremes, small or large, result in lower critical Reynolds numbers. They explained that small particles adjust almost immediately to the fluid velocity due to fast relaxation time; therefore, they only play the role of increasing the total density of the system and thus lowering the critical Reynolds number. However, at large values of $S$, the heavy and large particles are not able to interact with the fluid, thus resulting in critical Reynolds numbers similar to that of the clean flow. They also discovered that intermediate values of the dimensionless relaxation time yield the largest increase in modal stability, which is explained by the significant energy losses observed at these values.

In a numerical study, \citet{klinkenberg2011numerical} observed that the threshold energy for transition to turbulence initiated by two types of disturbances, streamwise vortices, and oblique waves, is increased by the introduction of particles. Subsequently, \citet{klinkenberg2013numerical} showed that the laminar-turbulent transition might be facilitated by the presence of particles at low number density. Additionally, they discussed the effects of stokes drag, added mass, and fluid acceleration on the modal and non-modal stabilities of mono-sectional particle-laden flows \citep{klinkenberg2014linear}. For modal instability, they suggested that particles lighter than the carrier fluid decrease the critical Reynolds number due to the fluid acceleration term.  In contrast, particles increase the critical Reynolds number when added mass and Stokes drag are considered instead of the fluid acceleration term. \citet{boronin2014modal} studied the stability of dusty-gas boundary layer flows with a non-uniform distribution of particles in the form of a localized dust layer. They found that the dusty gas flow's critical Reynolds number can be either substantially larger or smaller than that of the clean gas flow, depending on the position and width of the dust layer. Moreover, \citet{boronin2018effect} investigated the effects of gravity-induced particle settling and found that a uniform distribution of small-inertia particles in the main flow has a stabilizing effect in a vertical plane channel.   

Multiphase flow stability characteristics are complex and depend on various parameters introduced by considering a second phase. For example, \cite{schmidt2021global} developed a method for computing linear global modes of immiscible two-phase flows, considering surface tension, density, and viscosity difference between different phases. They showed that a surface-tension-induced destabilization of plane wakes could lead to oscillations depending on the flow's $We$ number. 

Due to complexities in such problems, monodispersed solid particles are usually assumed in theoretical hydrodynamic stability studies, with no particle interactions.
In realistic applications, however, droplets, bubbles, and solid particles may continuously change their shape and size due to vaporization, condensation, coalescence or breakup, and the different forces exerted by the flow~\citep{maxey1983equation}. During these processes, interphase and inter-particle heat and mass transfer may alter the stability characteristics of the carrier flow. 

One example is clustering instability, which occurs in laminar~\citep{Bellan01, bellan1991dynamics, Katoshevski07, sazhin2008particle, greenberg2011influence, greenberg2012spray} as well as in turbulent flows~\citep{falkovich2002acceleration, chen2006turbulent}. 
The mechanisms leading to clustering may involve particle interactions.
Friction and inelastic collisions of particles act as sinks of particle kinetic energy, which may lead to clustering instability~\citep{fullmer2017clustering}.
Non-uniformities in particle distribution may result in a transition to turbulence in flows laden with solid particles subjected to radiative heating; 
~\cite{zamansky2014radiation} proposed a mechanism for this instability and showed, using a direct numerical simulation, how heat transfer, buoyancy, and gravity lead to a feedback loop, which further enhances the preferential concentration and leads to flow instability.

The computation of interaction processes in dispersed multiphase flows is a very demanding task.
Multiphase flows display stability characteristics that differ from single-phase flows and flow structures might be further complicated with the introduction of evaporating droplets and the heat transfer process. Problems involving dispersed flows of evaporating fuel droplets are frequently encountered in the studies of combustion stability. The stability of flows is also affected by large recirculation zones and vortical structures in reacting and non-reacting flows~\citep{dagan2016, taamallah2019helical, chakroun2019flamelet, DAGAN2019368, dagan2021settling, avni2022dynamics}. We have recently studied the interactions of such unstable flows, which shows the overall destabilizing effect of evaporating sprays~\citep{dagan2015dynamics}.
In many practical applications of dispersed multiphase flows, the enormous number of particles and particle interactions further increase the complexity, and tracking each particle in a Lagrangian framework requires extreme computational resources that are currently unrealistic. 
Turbulent simulations are therefore very limited for the purpose of developing general stability models for dispersed multiphase flows.
On the other hand, a proper \textit{generalized} hydrodynamic stability theory that takes into account the particle interactions in such flows is currently not available. 

The objective of the present study is thus to develop a generalized hydrodynamic stability theory for interacting particle-laden flows, which may generally involve particle interactions, evaporation, coalescence, and condensation.

The extension of current hydrodynamic stability theories that take such interactions into account should be based upon mathematical formulations that have the ability to resolve interacting polydisperse particles. Thus, the description of the particulate matter should be coupled to the carrier flow through proper interphase mass transfer terms~\citep{Greenberg93}, and yet simple enough to solve analytically. 
A natural framework for this purpose is the sectional approach~\citep{tambour1984vaporization, Greenberg93}.
In the sectional approach, the particle size distribution is divided into size sections, and an averaging technique is applied within each section, whereby the sectional conservation equations are derived.   
More recently, this methodology was used by several authors to study \citep{kah2010eulerian} and used in similarity solutions for reacting and non-reacting monodisperse and polydisperse volatile droplets~\citep{dagan2017similarity, DAGAN2019368}, and the dynamics of reacting three-phase particles~\citep{dagan2018flame}.

It should be mentioned that several studies~\citep{de2009eulerian, de2009eulerian2,vie2015anisotropic} showed the limitation of the sectional approach in reproducing the dynamics of inertial enough particles correctly. Thus, the sectional approach is valid for droplets and particles with relatively low Stokes numbers.
Despite the aforementioned limitations, extensive use has been made of the sectional approach in a variety of polydisperse sprays in both non-reacting and reacting flows. This approach was shown to produce an excellent agreement with experimental data for different spray systems~\citep{Tambour85, Greenberg93, Katoshevski93, Katoshevski95, massot1998spray, Greenberg07}.

Notably, Lauren et al. introduced the multi-fluid method~\citep{laurent2001multi, laurent2004eulerian}, in which the mean velocity difference of the droplets is taken into account. This becomes important in coalescence processes in dense particle flows. 
~\cite{fox2008numerical} suggested an alternative to the Lagrangian approaches and applied the direct quadrature method of moments~\citep{marchisio2005solution} and the multi-fluid method in a numerical simulation. They showed that both Eulerian methods adequately describe the coalescence of a polydisperse evaporating spray. 
However, in the present study, we shall use a linearized variation of a discrete sectional approach for dilute particle-laden flows, for which the linear stability equations can be written in a modal form.

The mathematical modelling and formulations are presented in section \ref{sec:mathmodelling} and introduce particle-laden flow configurations and particle loadings in section \ref{sec:configurations}. Validations and verifications of the new method will be presented in section \ref{sec:results}, followed by results and discussions for monodisperse and polydisperse particle-laden channel flow.

\section{Mathematical Model}
\label{sec:mathmodelling}
A generalized mathematical formulation to investigate the hydrodynamic stability of polydisperse particle-laden shear flows is developed in this section. The particle-laden flow is described using a discrete sectional approach, introduced in section \ref{sec:intro}, where inter-phase and inter-particle interactions are modelled as source terms in both the carrier-flow and particulate matter equations. 
In the sectional approach, particles are divided into $j=1,2,...,N_s$ discrete size sections and transport equations of the particles' number density and momentum are written separately for each unique section $j$.

We consider an incompressible, viscous flow. Gravity or other body forces acting upon the flow or particles are not accounted for in the current representation, hence neglecting settling velocity for particles, although these terms may be added to our formulation without the loss of generality in future studies. Under these assumptions, the governing equations are presented as follows.
\subsection{Carrier-flow Equations}
 To generally model the possible effect of phase change on the stability characteristic of the flow, we consider the mass contribution in the carrier-flow gas phase from the sectional particulate phase. One example of such mass exchange is the evaporation of liquid droplets of different sizes. The sectional approach is applied to represent such mass transfer between sectional particles and between particles and the carrier flow. Note that the particle loading in the carrier flow is considered to be diluted enough such that the additional vapor mass concentration remains low.
The continuity equation can then be written as:
\begin{equation}
\frac{\partial u}{\partial x}+\frac{\partial v}{\partial y}+\frac{\partial w}{\partial z}=S_v~, 
\label{eq:gascontori}
\end{equation}
where $u,v$ and $w$ are the axial, vertical  and span-wise gaseous velocity components, respectively.

For the sake of generality, the equations are written here with the additional vapor production source term, $S_v$, that may account for additional gaseous mass flow rates from droplets to the carrier flow due to phase change, whereas no source term would be considered if the particulate matter consists of solid particles. 

$S_v$ is expressed as:
	\begin{equation}
S_v=-\sum _{j=1} ^{N_s} -C_j Q_{j}+B_{j,j+1} Q_{j+1}~.
\label{eq:Sv}
\end{equation}
Here, $N_s$ is the total number of sections, $C_j$ is the rate of particle leaving the $j$-th size section (due to evaporation etc.) and $B_{j,j+1}$ is the rate of particle transition from  section $j+1$ to section $j$. It should be noted that other interphase effects such as condensation of particles may also be explored using this formulation, although it will not be considered in the present study.
$Q_j$ \citep{saffman1962stability} is a dimensionless variable representing the mass  of the $j$-th section particulate phase per unit volume normalized by the host flow density, 
\begin{equation}
Q_j=\frac{N_j\cdot \frac{4}{3}\rho_p \pi a_j^3}{\rho_f}~,
\label{eq:Qjtilde}
\end{equation}
with $N_j$ and $a_j$ being the number density (number of particles per unit volume) and radius of the $j$-th section particles, respectively. $\rho_p$ and $\rho_f$ are the density of particle and carrier-flow, respectively.


Spherical particles affect the carrier-flow momentum in the form of viscosity-induced drag. The exchange of momentum is modelled by their relative velocity and the particles' relaxation time $\tau$. The corresponding \textit{sectional} representation of the Navier-Stokes (NS) equation for the carrier-flow takes the form:
\begin{equation}
\frac{\partial \va*{u}}{\partial t}+\va*{u} \cdot \nabla\va*{u}=-\frac{1}{\rho_f} \nabla p+\nu \nabla^2 \va*{u}+\sum^{N_s}_{j=1}\tau_j^{-1} Q_j(\va*{u}_j-\va*{u})~,
\label{eq:gasmomori}
\end{equation}
$p$ is pressure, $\nu$ is the flow kinematic viscosity and $\tau_j$ is the relaxation time of the particles in the $j$-th size section, which may be written explicitly as $\tau_j=\frac{\rho_p d_j^2}{18 \mu}$. $d_j$ denotes the mean value of the $j$-th section particles diameter. $\vec{u}_j$ is the velocity vector of $j$-th section particles, and the linear drag force is assumed.

\subsection{Disperse phase Equations}
In the sectional approach, particles are divided into $j=1,2,...,N_s$ discrete size sections. Transport equations of particle mass fraction and momentum are written separately for each unique section $j$. The dispersed phase balance equation is written in terms of the sectional mass fraction $Q_j$:
\begin{equation}
\frac{\partial Q_j}{\partial t}+u_j \frac{\partial Q_j}{\partial x}+v_j\frac{\partial Q_j}{\partial y}+w_j \frac{\partial Q_j}{\partial z}+Q_j\left(\frac{\partial u_j}{\partial x}+\frac{\partial v_j}{\partial y}+\frac{\partial w_j}{\partial z}\right)=S_j=-C_j Q_{j}+B_{j,j+1} Q_{j+1}
\end{equation}
Here, $S_j$ is the rate of net mass exchange of particles in section $j$ - evaporating droplet mass leaving the $j$-th section at rate $C_j$ and additional mass joining the $j$-th section from the $(j+1)$-th section at rate $B_{j,j+1}$. 
The sectional representation of particle momentum is:
\begin{equation}
Q_j \left(	\frac{\partial \va*{u}_j}{\partial t}+\va*{u}_j \cdot \nabla\va*{u}_j\right)+\va*{u_j}\left(\underbrace{	\frac{\partial Q_j}{\partial t}+\va*{u}_j \cdot \nabla Q_j+Q_j \nabla \cdot\va*{u}_j}_{=S_j} \right)=Q_j \tau_j^{-1}(\va*{u}-\va*{u}_j)+\va*{S}^{LM}_{j}
\end{equation}
\begin{equation}
Q_j \left(	\frac{\partial \va*{u}_j}{\partial t}+\va*{u}_j \cdot \nabla\va*{u}_j\right)=Q_j \tau_j^{-1}(\va*{u}-\va*{u}_j)+\left(\va*{S}^{LM}_{j}-\va*{u}_j S_j\right)~,
\label{eq:parmomori}
\end{equation}
where $\vec{S}^{LM}_{j}$ is the linear momentum exchange between neighbouring size section particles \citep{tambour1993derivation,katoshevski1993theoretical},

\begin{equation}
  \va*{S}^{LM}_{j}=-C_j Q_{j}\va*{u}_j+B_{j,j+1} Q_{j+1}\va*{u}_{j+1}.
\end{equation}
The mass transfer term $S_j$, and the linear momentum transfer $\vec{{S}_{j}}^{LM}$, are assumed to occur exclusively between two neighbouring sections.

\subsection{Dimensionless Formulation}
The equations of the previous sections describe the two-way coupled, three-dimensional particle-laden flow. In this study, the stability equations of two-dimensional polydisperse particle-laden flows will be derived. 
A two-dimensional formulation is considered henceforth. Using a characteristic length scale of the channel $H$ and the centre-line velocity $U_0$, we define the following dimensionless variables:
\begin{equation}
\begin{aligned}
&\tilde{u}=\frac{u}{U_0},\hspace{0.1in} \tilde{v}=\frac{v}{U_0},\hspace{0.1in} 
\tilde{u}_j=\frac{u_j}{U_0},\hspace{0.1in} \tilde{v}_j=\frac{v_j}{U_0},\hspace{0.1in} \\
&\tilde{x}=\frac{x}{H},\hspace{0.1in} \tilde{y}=\frac{y}{H},\hspace{0.1in}   \tilde{t}=\frac{t}{H/U_0},\hspace{0.1in}
\tilde{p}=\frac{P}{\rho_f U_0^2},\hspace{0.1in} \tilde{C}_j=\frac{C_j}{U_0/H}, \hspace{0.1in} \tilde{B}_{j,j+1}=\frac{B_{j,j+1}}{U_0/H}
\end{aligned}
\end{equation}
 Normalization of the dimensional evaporation coefficient rates $C_j [s^{-1}]$ and $B_{j,j+1} [s^{-1}]$ are denoted as $\tilde{C}_j$ and $\tilde{B}_{j,j+1}$. The flow Reynolds number is defined as $Re=\rho_f U_0 H/\mu$. The flow time-scale, $\tau_f$ and particle time scale $\tau_p$, determine a Stokes number $St\equiv\tau_p/\tau_f$. Here, particles of different size sections are distinguished by their \textit{sectional} Stokes number, $St_j$. 
 Using the aforementioned definitions (and omitting tilde notations for convenience), the dimensionless, two-dimensional equations are:  
\begin{equation}
\frac{\partial u}{\partial x}+\frac{\partial v}{\partial y}=-\sum^{N_s}_{j=1} -C_j Q_{j}+B_{j,j+1} Q_{j+1} 
\label{eq:gascontnd}
\end{equation}
\begin{equation}
\frac{\partial u}{\partial t} +u\frac{\partial u}{\partial x}+v\frac{\partial u}{\partial y}=-\frac{\partial p}{\partial x}+\frac{1}{Re}\nabla^2 u+\sum_{j=1}^{N_s} \frac{1}{St_j} Q_j (u_j-u)
\end{equation}
\begin{equation}
\frac{\partial v}{\partial t} +u\frac{\partial v}{\partial x}+v\frac{\partial v}{\partial y}=-\frac{\partial p}{\partial y}+\frac{1}{Re}\nabla^2 v+\sum_{j=1}^{N_s} \frac{1}{St_j} Q_j (v_j-v)
\end{equation}
\begin{equation}
\frac{\partial Q_j}{\partial t}+u_j \frac{\partial Q_j}{\partial x}+v_j\frac{\partial Q_j}{\partial y}+Q_j\left(\frac{\partial u_j}{\partial x}+\frac{\partial v_j}{\partial y}\right)= -C_j Q_{j}+B_{j,j+1} Q_{j+1}
\end{equation}
\begin{equation}
Q_j \left(	\frac{\partial u_j}{\partial t}+u_j \frac{\partial u_j}{\partial x}+v_j \frac{\partial u_j}{\partial y}\right)=\frac{Q_j}{St_j} (u-u_j)+ B_{j,j+1} Q_{j+1} \left( u_{j+1}-u_j\right)
\end{equation}
\begin{equation}
Q_j \left(	\frac{\partial v_j}{\partial t}+u_j \frac{\partial v_j}{\partial x}+v_j \frac{\partial v_j}{\partial y}\right)=\frac{Q_j}{St_j} (v-v_j)+ B_{j,j+1} Q_{j+1} \left( v_{j+1}-v_j\right)
\label{eq:parvjnd}
\end{equation}

\subsection{Linear Temporal Stability}
For linear temporal stability analysis, a steady, two-dimensional parallel base-flow is assumed, superimposed by perturbations in the streamwise $\hat{x}$ and transverse $\hat{z}$ directions. To formulate the modal linear stability equations, we shall decompose the carrier flow and particulate matter variables into a mean part and an infinitesimally small perturbation part, denoted here by $()^\prime$.
\begin{equation}
u=U+u^\prime; \hspace{0.1in} v=V+v^\prime; \hspace{0.1in} p=P+p^\prime; \hspace{0.1in} u_j=U_j+u_j^\prime; \hspace{0.1in} v_j=V_j+v_j^\prime; \hspace{0.1in} Q_j=Q_{0,j}+q_j^\prime; \hspace{0.1in} 
\end{equation}
The above base flow variables superposed with small perturbations are substituted into the dimensionless equations \ref{eq:gascontnd}-\ref{eq:parvjnd} and higher-order non-linear terms are omitted. 
 The following formulations have already been simplified by substituting the parallel plane Poiseuille base flow profile, assuming that the streamwise velocity $U$ is a function of the coordinate $y$ only and that the vertical base flow velocity $V$ is zero everywhere.
\begin{equation}
    U=U(y), ~~ V=0.
\end{equation}
The above assumptions also apply to the particulate matter base state profile, explicitly,
\begin{equation}
   U_j=U_j(y), ~~ V_j=0 ~~ \text{for} ~~ j=1,2,\dots, N_s.
\end{equation}
Therefore, the six dominant equations can be written as:
\begin{equation}
\frac{\partial u^\prime}{\partial x}+\frac{\partial v^\prime}{\partial y}=-\sum _{j=1} -C_j q^\prime_{j}+B_{j,j+1} q^\prime_{j+1} 
\label{eq:gascontnd}
\end{equation}
\begin{equation}
\frac{\partial u^\prime}{\partial t} +U\frac{\partial u^\prime}{\partial x}+v^\prime\frac{\partial U}{\partial y}=-\frac{\partial p^\prime}{\partial x}+\frac{1}{Re}\nabla^2 u^\prime+\sum_{j=1}^{N_s} \frac{1}{St_j}\left[ Q_{0,j} (u^\prime_j-u^\prime)+q_j^\prime(U_j-U)\right] 
\label{eq:pert_u}
\end{equation}
\begin{equation}
\frac{\partial v^\prime}{\partial t} +U\frac{\partial v^\prime}{\partial x}=-\frac{\partial p^\prime}{\partial y}+\frac{1}{Re}\nabla^2 v^\prime+\sum_{j=1}^{N_s} \frac{1}{St_j}\left[   Q_{0,j} (v^\prime_j-v^\prime)+q_j^\prime(V_j-V)\right] 
\end{equation}
\begin{equation}
\frac{\partial q^\prime_j}{\partial t}+U_j \frac{\partial q^\prime_j}{\partial x}+u_j^\prime \frac{\partial Q_{0,j}}{\partial x}+v_j^\prime \frac{\partial Q_{0,j}}{\partial y}+Q_{0,j}\left(\frac{\partial u^\prime_j}{\partial x}+\frac{\partial v^\prime_j}{\partial y}\right)= -C_j q^\prime_{j}+B_{j,j+1} q^\prime_{j+1}
\end{equation}
\begin{equation}
\begin{gathered}
Q_{0,j} \left(	\frac{\partial u^\prime_j}{\partial t}+U_j \frac{\partial u^\prime_j}{\partial x}+v^\prime_j \frac{\partial U_j}{\partial y}\right)= \frac{1}{St_j}\left[ Q_{0,j} (u^\prime-u^\prime_j)+q^\prime_{j}\left( U-U_j\right)\right] \\
+B_{j,j+1}\left[  Q_{0,j+1} \left( u^\prime_{j+1}-u^\prime_j\right)+q^\prime_{j+1}(U_{j+1}-U_j)\right]
\end{gathered}
\end{equation}
\begin{equation}
\begin{gathered}
Q_{0,j} \left(	\frac{\partial v^\prime_j}{\partial t}+U_j \frac{\partial v^\prime_j}{\partial x}\right)=\frac{1}{St_j}\left[ Q_{0,j} (v^\prime-v^\prime_j)\right]+B_{j,j+1}\left[  Q_{0,j+1} \left( v^\prime_{j+1}-v^\prime_j\right)\right]
\end{gathered}
\label{eq:pert_vj}
\end{equation}

\subsection{Normal Mode Formulation}
\label{section:normalmodes}
The set of coupled linear equations may be represented in a general normal modes form, where perturbations are considered to be wave-like functions propagating in time and space. Perturbations in the polydisperse channel flow configuration are defined as
\begin{equation}
\{u^\prime,v^\prime,p^\prime,q^\prime_j,u^\prime_j,v^\prime_j\}=\{F(y), iG(y), P(y), M_j(y), F_j(y), iG_j(y) \} e^{i(\alpha x+\beta z-\omega t)}~,
\label{eq:eigfunc}
\end{equation}
where $\alpha$ and $\beta$ are wavenumbers in the $\hat{x}$ and $\hat{z}$ directions, respectively, and $\omega$ is the temporal frequency of the disturbances. In this study we focus on examining the temporal instabilities of two-dimensional particle-laden flows and therefore assume only streamwise propagating perturbations, i.e. $\beta=0$. For a clean shear flow, a two-dimensional perturbation wave will result in the least stable growth according to Squire's theorem \citep{squire1933stability}.  $F, G, P, M_j, F_j, G_j$ are the corresponding eigenmodes, where particle eigenmodes are denoted by the section number $j$. The derivation operators become:
\[
\begin{split}
\frac{\partial}{\partial t}=-i\omega \hspace{0.3in}   \frac{\partial}{\partial x}= i\alpha \hspace{0.3in}  \frac{\partial}{\partial y}= D \hspace{0.3in}   \nabla^2=D^2-\alpha^2
\end{split}
\]
where $D$ is defined as derivative with respect to vertical coordinate $y$. Substituting the differential operators defined above into the linearized equations, the normal mode formulation may be written as follows.
\begin{equation}
i\alpha F+i DG+\sum_{j=1}^{N_s}-C_j M_j+B_{j,j+1}M_{j+1}=0
\label{eq:efgasmass_ch}
\end{equation}
\begin{equation}
	\left[  i\alpha U-\frac{1}{Re}\left(D^2-\alpha^2\right) +\sum^{N_s}_{j=1} \frac{Q_{0,j}}{St_j}\right]F+i\frac{\partial U}{\partial y}G+i\alpha P -\sum^{N_s}_{j=1} \frac{Q_{0,j}}{St_j} F_j-\sum^{N_s}_{j=1} \frac{U_j-U}{St_j} M_j=i \omega F
\end{equation}
\begin{equation}
	\left[ -\alpha U -\frac{i}{Re}\left(D^2-\alpha^2\right)+i\sum^{N_s}_{j=1} \frac{Q_{0,j}}{St_j}\right] G +DP-i\sum^{N_s}_{j=1} \frac{Q_{0,j}}{St_j} G_j-\sum^{N_s}_{j=1} \frac{V_j-V}{St_j} M_j=-\omega G
\end{equation}
\begin{equation}
\begin{aligned}
\left( i\alpha Q_{0,j}+\frac{\partial Q_{0,j}}{\partial x}\right) F_j+\left( i Q_{0,j}D+i\frac{\partial Q_{0,j}}{\partial y}\right) G_j+&\left( \frac{\partial U_j}{\partial x}+\frac{\partial V_j}{\partial y}+i\alpha U_j+V_j D+C_j\right) M_j\\
&-B_{j,j+1}M_{j+1}=i\omega M_j
\end{aligned}
\label{eq:efparmass_ch}
\end{equation}
\begin{equation}
\begin{aligned}
&-\frac{Q_{0,j}}{St_j} F+\left[ Q_{0,j}\left( i\alpha U_j +\frac{\partial U_j}{\partial x}+V_j D\right) +\frac{Q_{0,j}}{St_j} +B_{j,j+1}Q_{0,j+1}\right] F_j+i Q_{0,j}\frac{\partial U_j}{\partial y} G_j\\
& \left( U_j\frac{\partial U_j}{\partial x}+V_j \frac{\partial U_j}{\partial y}-\frac{U-U_j}{St_j}\right) M_j-B_{j,j+1}Q_{0,j+1}F_{j+1}-B_{j,j+1}\left(U_{j+1}-U_j \right) M_{j+1}=i\omega Q_{0,j}F_j
\end{aligned}
\end{equation}
\begin{equation}
\begin{aligned}
&-i\frac{Q_{0,j}}{St_j} G+ Q_{0,j}\frac{\partial V_j}{\partial x} F_j+\left[ Q_{0,j}\left( -\alpha U_j +iV_j D+i\frac{\partial V_j}{\partial y}\right) +i\frac{Q_{0,j}}{St_j} +iB_{j,j+1}Q_{0,j+1}\right] G_j\\
& \left( U_j\frac{\partial V_j}{\partial x}+V_j \frac{\partial V_j}{\partial y}-\frac{V-V_j}{St_j}\right) M_j-iB_{j,j+1}Q_{0,j+1}G_{j+1}-B_{j,j+1}\left(V_{j+1}-V_j \right) M_{j+1}=-\omega Q_{0,j}G_j
\end{aligned}
\label{eq:efparv}
\end{equation}
The linear system of equations may be arranged in the form of a generalized eigenvalue problem, $\mathbf{A X}=\omega \mathbf{B X}$ with $\omega$ being the eigenfrequencies and $\mathbf{X}$ being the eigenvector in equation \ref{eq:eigfunc}. Matrices $\mathbf{A}$ and $\mathbf{B}$ are presented in detail in appendix \ref{section:app}.

\section{Particle-laden Flow Configuration}
\label{sec:configurations}
\subsection{Base Flow Profile and Boundary Conditions}
\begin{figure}
	\centerline{\includegraphics[width=0.9\textwidth]{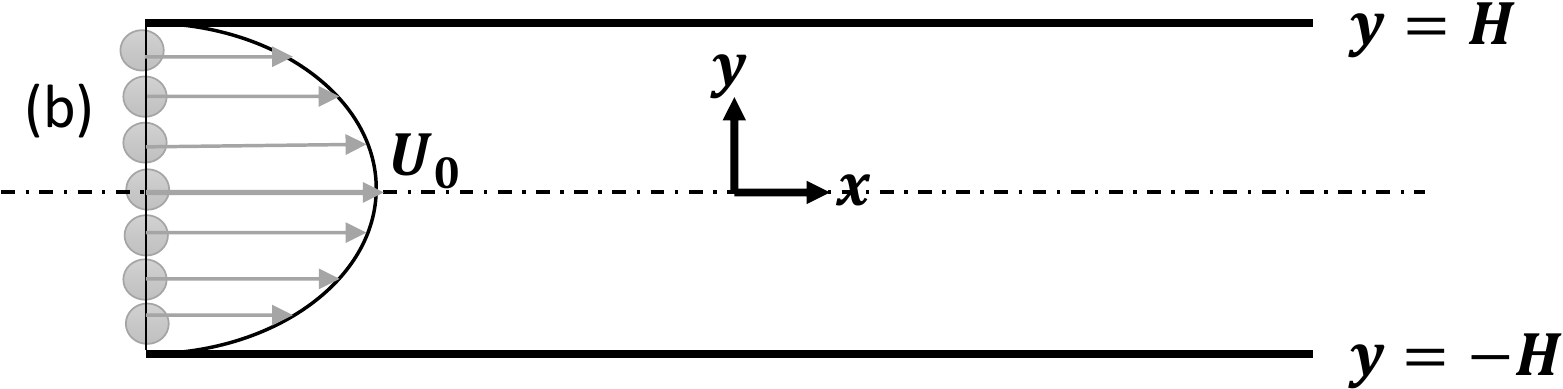}}
	\caption{Channel flow with a parabolic base flow profile $U(y)=U_0\left(1-(\frac{y}{H})^2\right), V(y)=0$. Reference parameters are taken as half the channel height, $H$ and the centre-line velocity $U_0$. Base-flow particles are uniformly distributed across the channel with the same profile as its carrier-flow, $U_j(y)=U(y)=U_0\left(1-(\frac{y}{H})^2\right), V_j(y)=V(y)=0$. }
	\label{fig:channels}
\end{figure}

Here, we consider channel flows laden with polydisperse disperse particulate matter. Note that the equations derived above are applicable to two-dimensional parallel particle-laden channel flows. Here, we specifically study the frequently encountered plane Poiseuille channel flow with the parabolic base-flow profile, $U(y)=U_0\left(1-(\frac{y}{H})^2\right)$, as shown in figure (\ref{fig:channels}).
$U_0$ is the base-flow centre-line velocity, and $H$ is half the channel height. The parallel base-flow assumption requires that the base-flow vertical velocity should be zero at any location, and thus, $V(y)=0$. All velocity components vanish at solid boundaries $y=\pm H$.
For the particulate matter, we assume the base-flow velocities $U_j, V_j$ are identical to those of the carrier flow-field. 
Boundary conditions are defined such that all perturbations vanish at solid boundaries,
\begin{equation}
u^\prime=0, \hspace{0.1in} v^\prime=0, \hspace{0.1in} p^\prime=0, \hspace{0.1in} u_j^\prime=0, \hspace{0.1in} v_j^\prime=0, \hspace{0.1in} q_j^\prime=0 \hspace{0.1in} \text{for} \ j=1,2,...,N_s \ \text{at} \ y=\pm H
\end{equation}

\subsection{Polydisperse Particle Loading}
In addition to studying the effect of monodisperse particle distribution, we aim to understand the influence of polydispersity on the stability of the flow. In particular, we explore the effect of adding fine dust or mist to coarse particles, where the averaged Stokes number of all particles is to be kept consistent between the mono - and polydisperse - systems.  

We consider polydisperse particles of two size sections with Stokes numbers $St^\ast$ and $St^\dagger$, where $St^\ast$ is much smaller than $St^\dagger$. 
We then compare the stability characteristics of monodisperse particle cases with Stokes number $St^{SMD}$ to that of the bi-sectional distribution. Monodisperse particles of size $St^\dagger$, which is chosen deliberately only slightly larger than $St^{SMD}$, display similar stability characteristics as those of $St^{SMD}$. Therefore, when fine particles with $St^\ast$ are added, the change in stability characteristics will mostly reflect the effects of fine particle addition to an existing monodisperse of coarse particles. Realization of this analysis is carried out by designing a bi-sectional disperse system that shares the same averaged Stokes number and the same total particulate loading as a monodisperse system.

We shall now define a Sauter mean diameter (SMD), denoted by $D_{32}$, to define a mean characteristic value for particle diameters of two sizes with their respective number densities. Consider a polydisperse distribution with two particles sizes ($j=1,2$ and $N_s=2$). All particles in the $j=1$ section have a diameter of $d_1$ and a number density fraction $n_1$. The remaining particles, of section $j=2$, all have have a diameter of $d_2$ and a number density fraction $n_2$. Throughout this study, properties of particles with diameters smaller than the reference particle will be denoted by $^{\ast}$, whereas properties of particles with diameters larger than the reference particle will be denoted by $^{\dagger}$. The particle size distribution is discrete in this case, and thus, according to its definition \citep{kowalczuk2016physical}, the SMD of this bi-sectional disperse can be calculated using:
\begin{equation}
D_{32}=\frac{(D^\ast)^3 n^\ast+(D^\dagger)^3 n^\dagger}{(D^\ast)^2 n^\ast+(D^\dagger)^2 n^\dagger} ~.
\label{eq:defSMD}
\end{equation}
Here, $D^\ast, n^\ast$ are the diameter and number density of the smaller sized particles, and $D^\dagger, n^\dagger$ are those of the larger sized particles in the following bi-sectional particle study.

When particle density, carrier-flow viscosity and flow time scale $\tau_f$ are kept constant, the particle Stokes number is solely determined by the particle diameter $St\propto D^2$. The relation may be expressed as:
\begin{equation}
\frac{D^\ast}{D^\dagger}=\sqrt{\frac{St^\ast}{St^\dagger}}
\end{equation}
The definition of SMD $d_{32}$ in equation (\ref{eq:defSMD}) yields an equation relating number densities with Stokes number ratios:
\begin{equation}
1=\frac{\left( \sqrt{\frac{St^\ast}{St^{SMD}}}\right) ^3 n^\ast+\left( \sqrt{\frac{St^\dagger}{St^{SMD}}}\right) ^3 n^\dagger}{\left( \sqrt{\frac{St^\ast}{St^{SMD}}}\right) ^2 n^\ast+\left( \sqrt{\frac{St^\dagger}{St^{SMD}}}\right) ^2 n^\dagger}~.
\label{eq:stratio}
\end{equation}
And since the polydisperse system consists of only two sections,
\begin{equation}
    n^\ast+n^\dagger=1~.
\end{equation}
Let us rewrite the above equation in terms of number densities $N^\ast=n^\ast \cdot N_{tot}$ and $N^\dagger=n^\dagger \cdot N_{tot}$ with $N_{tot}$ being the total number density of particles from two sections:
\begin{equation}
1=\frac{\left( \sqrt{\frac{St^\ast}{St^{SMD}}}\right) ^3 \frac{N^\ast}{N_{tot}}+\left( \sqrt{\frac{St^\dagger}{St^{SMD}}}\right) ^3 \frac{N^\dagger}{N_{tot}}}{\left( \sqrt{\frac{St^\ast}{St^{SMD}}}\right) ^2 \frac{N^\ast}{N_{tot}}+\left( \sqrt{\frac{St^\dagger}{St^{SMD}}}\right) ^2 \frac{N^\dagger}{N_{tot}}}=\frac{\left( \sqrt{\frac{St^\ast}{St^{SMD}}}\right) ^3 N^\ast+\left( \sqrt{\frac{St^\dagger}{St^{SMD}}}\right) ^3 N^\dagger}{\left( \sqrt{\frac{St^\ast}{St^{SMD}}}\right) ^2 N^\ast+\left( \sqrt{\frac{St^\dagger}{St^{SMD}}}\right) ^2 N^\dagger}
\label{eq:2secst}
\end{equation}
The total particulate matter mass fraction with respect to flow mass, $Q_{0,tot}$ is to be kept constant according to the definition in equation (\ref{eq:Qjtilde}). It is also required that
\begin{equation}
    Q_{0,tot}=\frac{\frac{1}{6}\rho_p \pi }{\rho_f}\left[ N^{\ast} (D^{\ast})^3+ N^{\dagger} (D^{\dagger})^ 3\right]
\label{eq:2secqtot}
\end{equation}

We are now able to express the bi-sectional polydisperse particle loading with a desired $St^{SMD}$ and two given particle sizes $D^\ast$, $D^\dagger$, and determine the required number densities $N^\ast$, $N^\dagger$ using equations (\ref{eq:2secst}) and (\ref{eq:2secqtot}). 
Through these constraints, bi-sectional disperse systems with double-sized particles of $St^\ast$ and $St^{\dagger}$ are expected to experience stability characteristics that are closely related to that of a monodisperse of $St^{SMD}$ since the choice of $St^{\dagger}$ depends on the desired $St^{SMD}$: $St^{\dagger}$ is slightly higher than $St^{SMD}$. 
Specifically, the monodisperse case is equivalent to the bi-sectional case composed of two same sections particles, i.e. $St^\ast=St^\dagger=St^{SMD}$.
However, the stability characteristics may also differ from those of $St^{SMD}$ monodisperse case when finer particles of $St^\ast$ are injected into the disperse systems. 
By comparing the stability characteristics of mono- and 2-section dispersion with the same averaged Stokes number, $St^{SMD}$, and the same total particle loading, we are able to investigate the influence of fine particles addition to coarse particle-laden flows.

Table \ref{table:cases} shows the mono-section and 2-section cases examined in the present study with particle mass fraction prescribed by their Stokes numbers, based on equation (\ref{eq:stratio}). Three different $St^{SMD}$'s representing different regions of examined Stokes numbers are chosen, namely $St^{SMD}=0.1$, $St^{SMD}=2$ and $St^{SMD}=10$. 
To observe the effect of adding fine particles to the mono-sectional cases and to construct 2-sectional cases with these $St^{SMD}$, fine particles are arbitrarily chosen to be of $St^{\ast}=0.01$.
Coarse particles are chosen to be close to the desired $St^{SMD}$'s, namely $St^{\dagger}=0.2$, $St^{\dagger}=2.1$ and $St^{\dagger}=10.1$.

\begin{table}
	\centering
	\begin{tabular}{ccccccccc}
		\hline
		& j=1      & j=2 & j=3      & j=4 & j=5     & j=6 & j=7      &  \\
		St     & 0.01     & 0.1 & 0.2      & 2   & 2.1     & 10  & 10.1     &   $St^{SMD}$         \\
		\hline
		case 1 & 0        & 1   & 0        & 0   & 0       & 0   & 0        & 0.1        \\
		case 2 & 0.119296 & 0   & 0.880704 & 0   & 0       & 0   & 0        & 0.1        \\
		case 3 & 0        & 0   & 0        & 1   & 0       & 0   & 0        & 2          \\
		case 4 & 0.00183  & 0   & 0        & 0   & 0.99817 & 0   & 0        & 2          \\
		case 5 & 0        & 0   & 0        & 0   & 0       & 1   & 0        & 10         \\
		case 6 & 0.000162 & 0   & 0        & 0   & 0       & 0   & 0.999838 & 10         \\ \hline
	\end{tabular}
\caption{Sectional mass fraction of particles in different bi-sectional cases.}
\label{table:cases}
\end{table}

\section{Results and discussion}
\label{sec:results}
\subsection{Validation and Verification for Mono-sectional Particle-Laden Flows}
Previous studies on hydrodynamic stability of parallel flows, including dispersed particulate matter, have extensively adopted the Orr-Sommerfeld equation solution for its simplicity. In this study, due to the complexity of interacting polydisperse particle-laden flow problems, the interactions between particles and the carrier-flow are resolved using linear stability of NS equations with additional source terms, following the sectional approach, as shown in equations (\ref{eq:gascontori})-(\ref{eq:parmomori}).

 We start by comparing the most unstable temporal growth rate $\omega_{im,max}$ of a particle-free (clean) plane Poiseuille flow calculated by both OS and NS equations with different collocation numbers to verify the results of our new method. The OS equation,
\begin{equation}
	\left[(-i \omega+i \alpha U)\left(\mathcal{D}^{2}-k^{2}\right)-i \alpha U^{\prime \prime}-\frac{1}{\operatorname{Re}}\left(\mathcal{D}^{2}-k^{2}\right)^{2}\right] \tilde{v}=0 ~,
\end{equation}
can thus be solved as an eigenvalue problem using the Chebyshev collocation method on the flow domain.  The resulting eigen-values indicate the stabilities of the particle-free parallel flows and is validated with previous results. 
Recall that according to our definition of the complex frequency $\omega$, a positive value of its imaginary part means a temporally growing perturbation leading to instability, while a negative value corresponds to a decaying perturbation.

It should be noted that for the polydisperse particle-laden NS equations, the solution of the full set of equations, including the particulate matter phase, is also solved here, although nullified for the case of clean channel flow. This, in turn, significantly increases the matrices size that may become ill-conditioned and thus require further verification.

\begin{figure}
	\centerline{\includegraphics[width=0.6\textwidth]{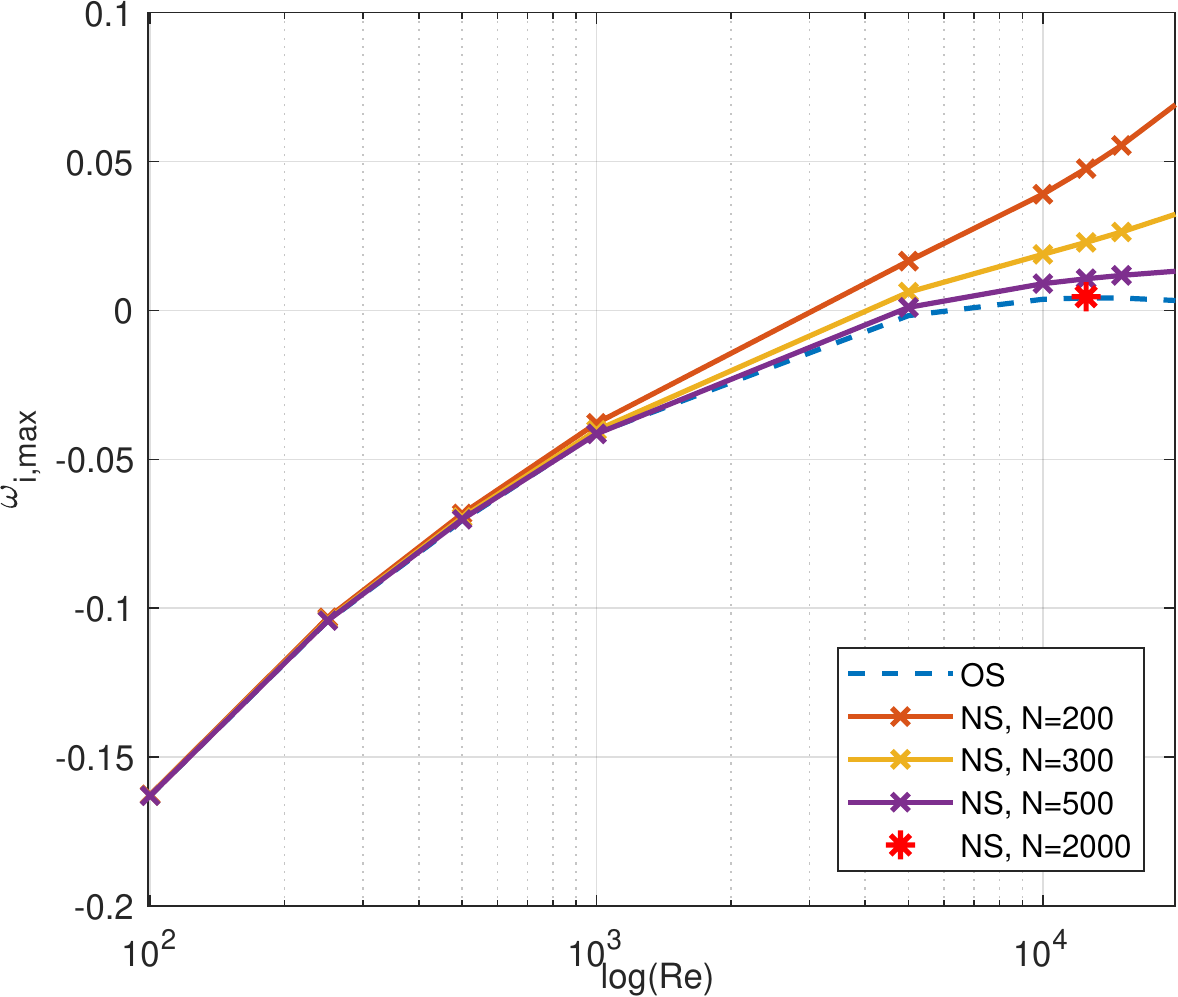}}
	\caption{The most unstable eigenvalues for the clean flow, extracted using the Orr-Sommerfeld equation, and the present study method with increasing collocation number $N=200, N=300, N=500$ across a wide range of Reynolds numbers and $N=2000$ for when $Re=12500$.  }
	\label{fig:NSOS}
\end{figure}

Results for the validation are shown in figure \ref{fig:NSOS}. We noticed a discrepancy between the two results that could be originated from the unstable nature of flows at high Reynolds numbers. We observe that increasing of Reynolds number leads to the growth of discrepancy, which requires a higher collocation number for convergence. For a high Reynolds number of $Re=12500$, and by using either $N=2000$ or $N=3000$ collocation points, we obtain satisfying results with respect to those obtained by OS. Henceforth, for such high Reynolds numbers, we shall use N=2000 for our analysis. Generally, for high $Re$, we obtained over-predictions of the growth rates compared to OS results. For example, at $Re=12500$ result using $N=2000$ yields an over-prediction of $3.91e-4$ compared to the OS result.

 In a typical stability spectrum of a perturbed particle-free Poiseuille flow, three branches of eigenvalues appear, namely $A$, $P$ and $S$ family \citep{drazin2004hydrodynamic}. Classification of eigenmodes into these three branches was discussed in detail by \citet{mack1976numerical}; eigenmodes on different branches represent instabilities that result from different physical mechanisms. In a recent study on temporal instability of a two-phase channel flow, \citet{kaffel2015eigenspectra} point out that $A$ branch eigenvalues are associated with the shear mode found in the classical hydrodynamic stability theory, while those on the $P$ branch are associated with the interfacial mode due to viscosity and density stratifications in two-fluid flows.
 
\begin{figure}
	\centerline{\includegraphics[width=0.6\textwidth]{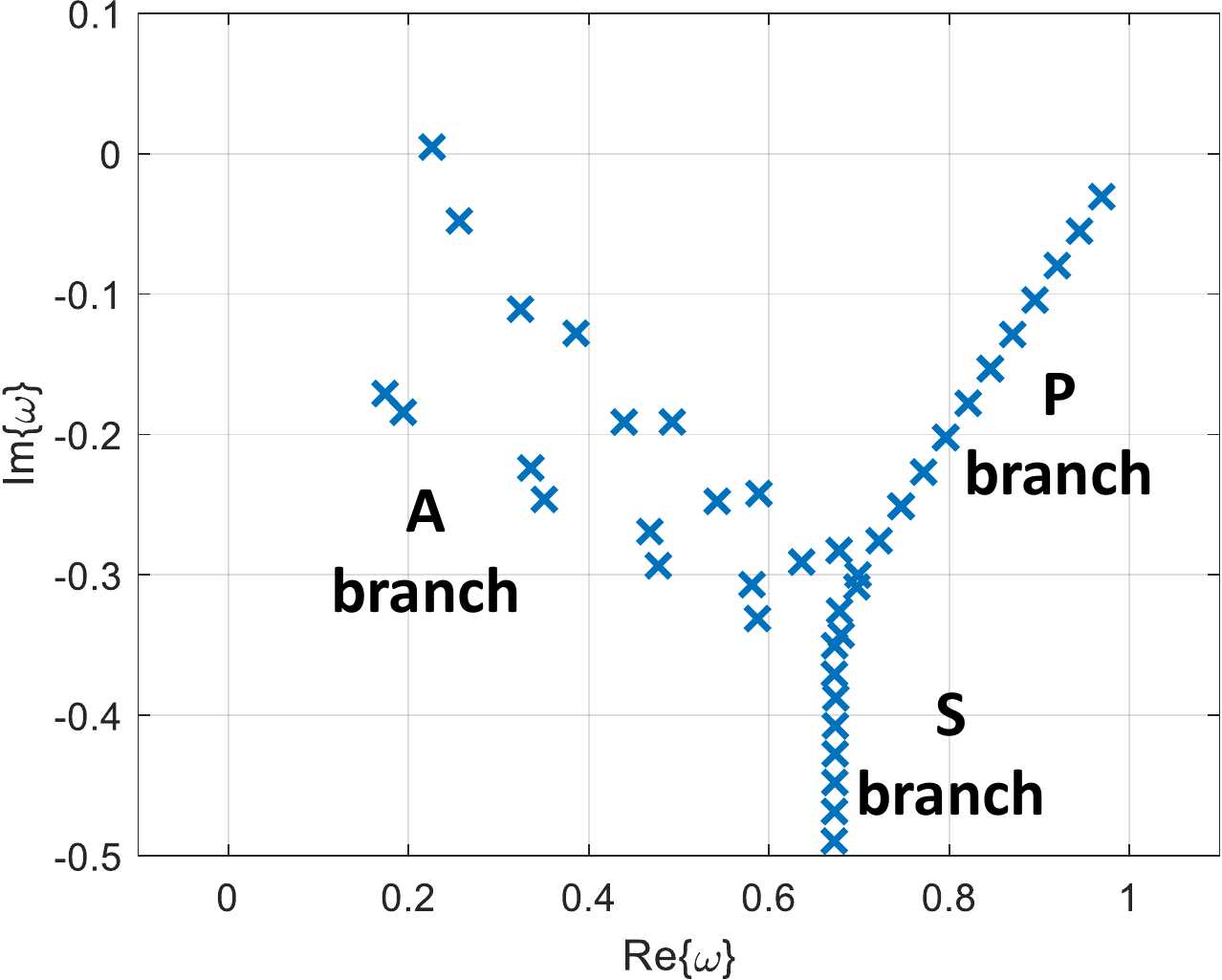}}
	\caption{Eigenvalue map for $St=0.001$ monodisperse channel flow at $Re=12500$. Three branches of eigenvalues, $A$, $P$ and $S$, are indicated.}
	\label{fig:APS}
\end{figure}

 Similar Y-shaped eigenvalue maps with these three branches are also observed here in monodisperse channel flows. Figure \ref{fig:APS} presents such an eigenvalue map for $St=0.001$ and $Re=12500$, where the $A$, $P$ and $S$ branches. In this case, and similar to the particle-free flow, the most unstable eigenvalue is located on the $A$ branch.
 
\begin{figure}
	\centering
	\begin{subfigure}[b]{0.45\textwidth}
		\centering
		\includegraphics[width=\textwidth]{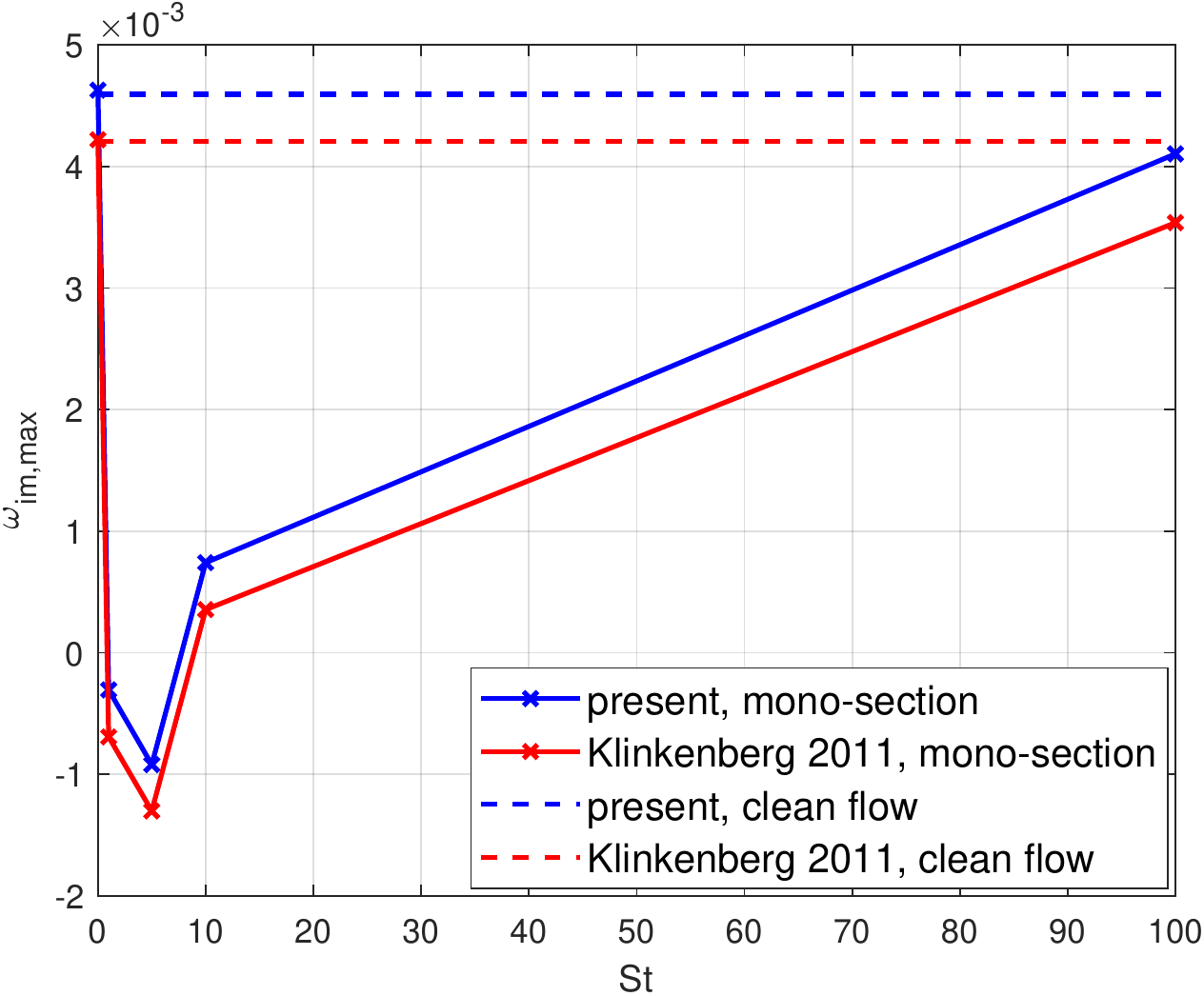}
		\caption{Collocation number $N=2000$}
		\label{fig:comparisona}
	\end{subfigure}
	\begin{subfigure}[b]{0.45\textwidth}
		\centering
		\includegraphics[width=\textwidth]{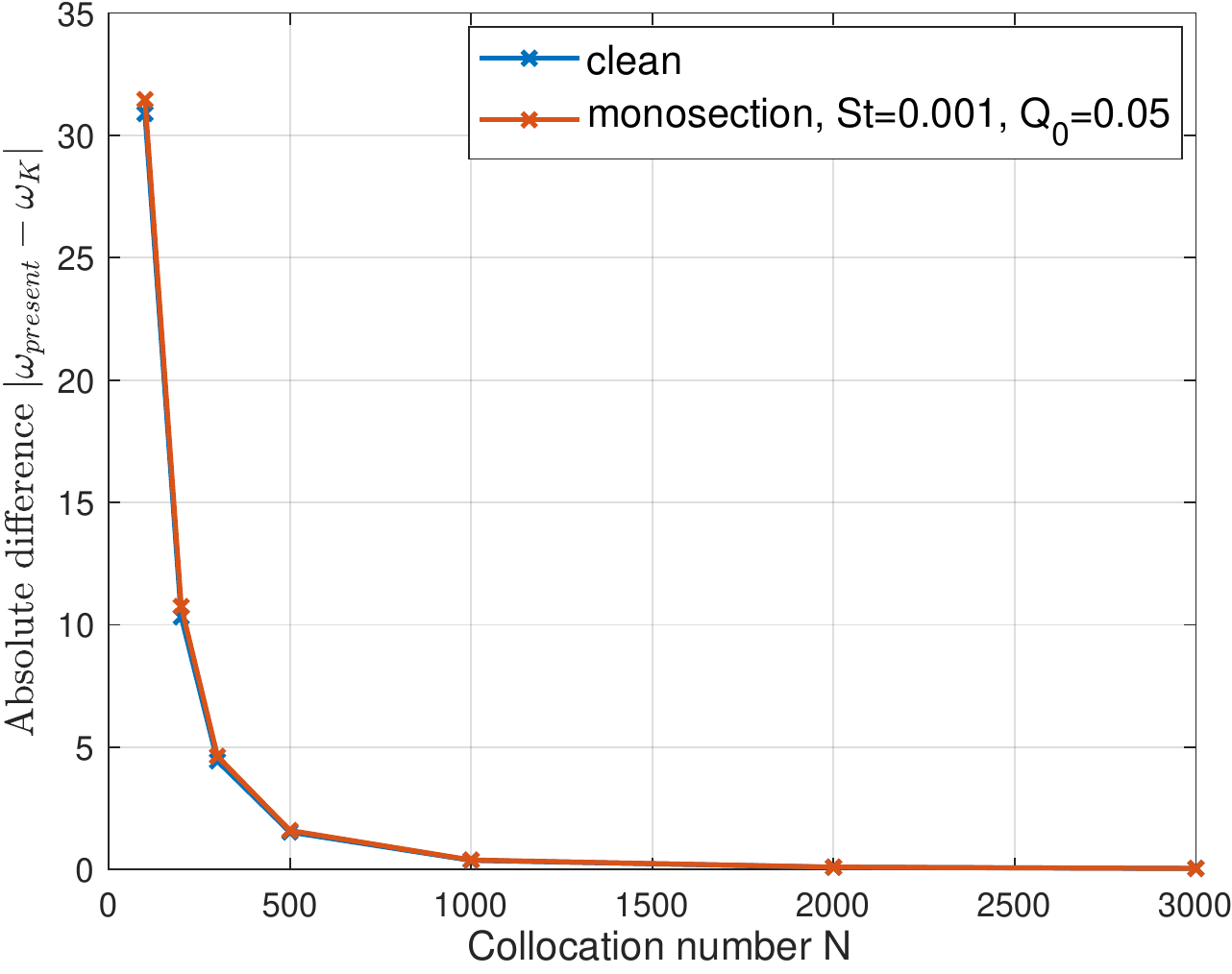}
		\caption{Absolute difference}
		\label{fig:comparisonb}
	\end{subfigure}
	\caption{(a) The most unstable eigenvalues of the $A$ family branch, calculated for $Re=12500$ and wave-number $\alpha=1$, with particle loading of $Q_{0,j}=0.05$ for monodisperse particle-laden channel flows. Solid lines indicate results of particle-laden flow with varying Stokes numbers of the present study, and of \cite{klinkenberg2011modal}, displaying the same trend with a nearly constant offset of $\omega_{im,max}$ for all Stokes numbers.
	Dashed lines indicate the results of a clean flow of both studies.
	 (b) The absolute difference as a function of the Chebyshev collocation number, compared with \cite{klinkenberg2011modal}. The blue marks indicate differences in the clean flow eigenvalues for $Re=12500$ and $\alpha=1$; red marks indicate differences in the monodisperse flow loaded with particles of $St=0.001$ and mass fraction $Q_{0,j}=0.05$.}
	\label{fig:comparison}
\end{figure}

Figure \ref{fig:comparison} compares the most unstable eigenvalues on the $A$ family branch obtained using our sectional linear stability (SLS) approach to the results reported by \cite{klinkenberg2011modal} for a monodisperse case. In their modal analysis, eigenvalues were derived from a modified OS equation for particle-laden flows. Results are shown for the case of $Re=12500, \alpha=1, Q_{0,j}=0.05$ and $St$ from 0.001 to $100$. The solid lines represent results of particle-laden cases, which are obtained here using 2000 collocations, whereas dashed lines indicate clean flow eigenvalues under the same conditions. 
  
Discrepancies between the two methods are approximately the same at different Stokes numbers and the clean flow, although the results of the most unstable growth rates are highly sensitive to the collocation number $N$ using the sectional stability approach.
  
  Figure \ref{fig:comparisonb} shows that the absolute differences between the results of present study and those presented in \cite{klinkenberg2011modal}, converge to zero as N increases. Therefore, the results to be discussed later are calculated using $N=2000$ collocations in the domain for computational efficiency.  \\

\begin{figure}
	\centerline{\includegraphics[width=0.6\textwidth]{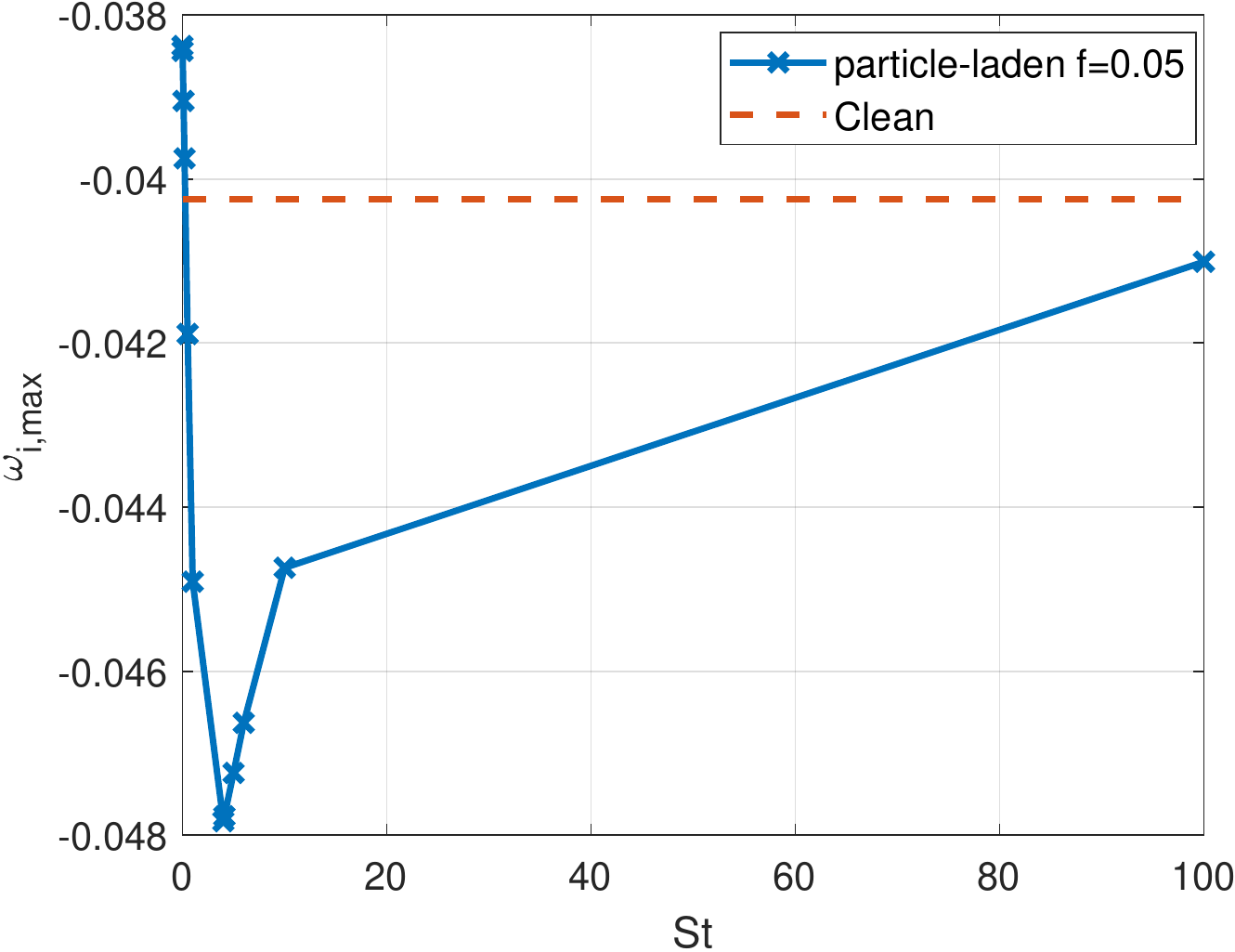}}
	\caption{The most unstable eigenvalues of the $A$ family branch calculated for a channel flow of $Re=1000$ and wave-number $\alpha=1$  with loading of $Q_{0,j}=0.05$ and different Stokes numbers of monodisperse particles. Addition of fine particles with $St<0.2$ results in lower perturbation decaying rates compared with the clean flow, while addition of larger particles  with $St\geq0.2$ result in faster perturbation decaying rates compared with the clean flow. The most stable case is obtained for particles addition at $St=4$.}
	\label{fig:lowRe}
\end{figure}

We proceed by analysing the stability of mono-sectional particle-laden flows for a lower Reynolds number of $Re=1000$, and compare the stability characteristics between low and high Reynolds numbers. Figure \ref{fig:lowRe} shows that for $Re=1000$, the stability has a similar trend as for $Re=12500$ with increasing Stokes numbers, as shown in figure \ref{fig:comparison}; the flow becomes less stable, compared to its clean state, with the addition of very fine particulate matters of $St\sim0.001 - 0.1$, whereas it becomes the most stable with the addition of particles with $St$ around unity and becomes less stable again with the increase of $St$. 
Note that based on our verification with the clean flow OS solution, we expect higher accuracy for this case as the Reynolds number, in this case, is lower.

\subsection{Stability of Polydisperse Channel Flows}

\begin{figure}
	\centerline{\includegraphics[width=1\textwidth]{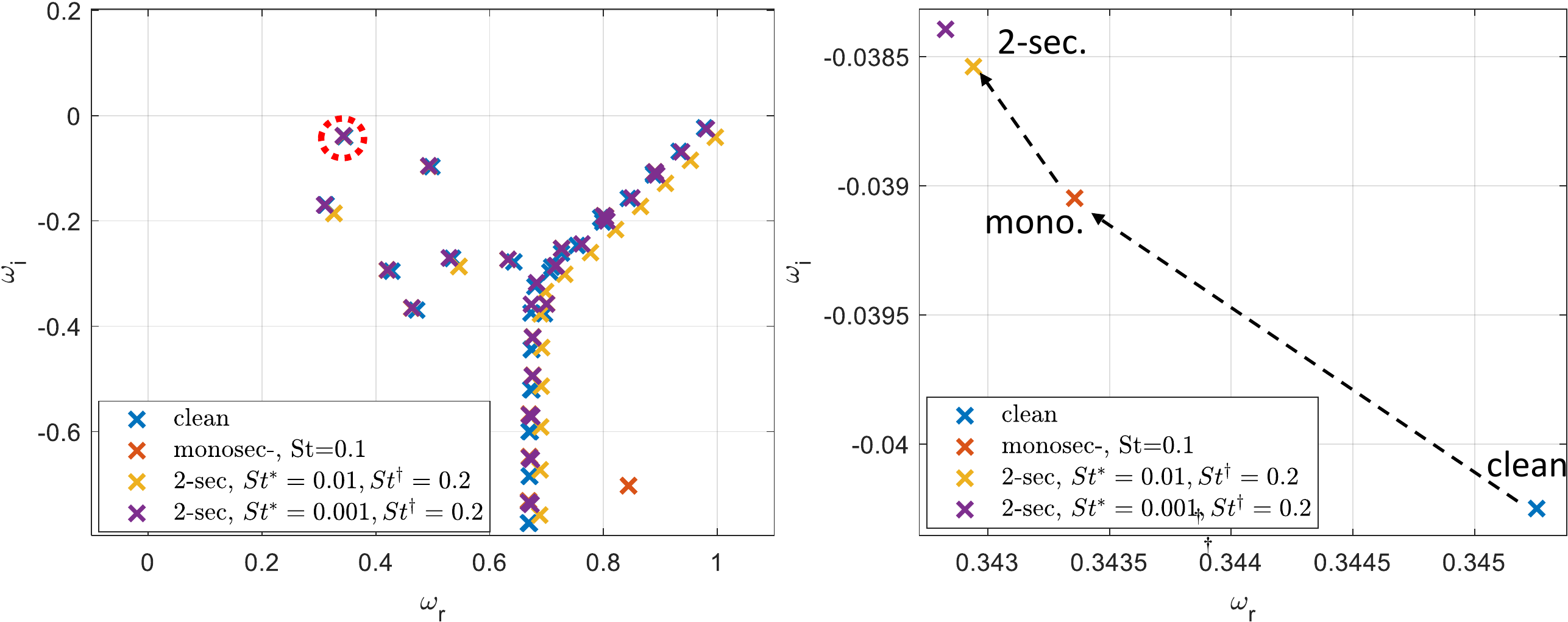}}
	\caption{Stability Spectra of $Re=1000$ clean flow , mono-sectional particle-laden flow with $St=0.1$, and bisectional particle-laden flow with combinations of: $St^\ast=0.01, St^\dagger=0.2$; $St^\ast=0.001, St^\dagger=0.2$. The four cases display similar shapes with a small shift to the lower-right for the two-sectional $St^\ast=0.01, St^\dagger=0.2$ case.  (b) a zoom-in of the most unstable mode of the $A$ family (dashed red circle of the left branch); for this mode, the addition of particles, either mono-sectional or bi-sectional, increases the growth rate and decreases the frequency of the clean flow.}
	\label{fig:yshape1Kst0.1}
\end{figure}

\begin{figure}
	\centerline{\includegraphics[width=1\textwidth]{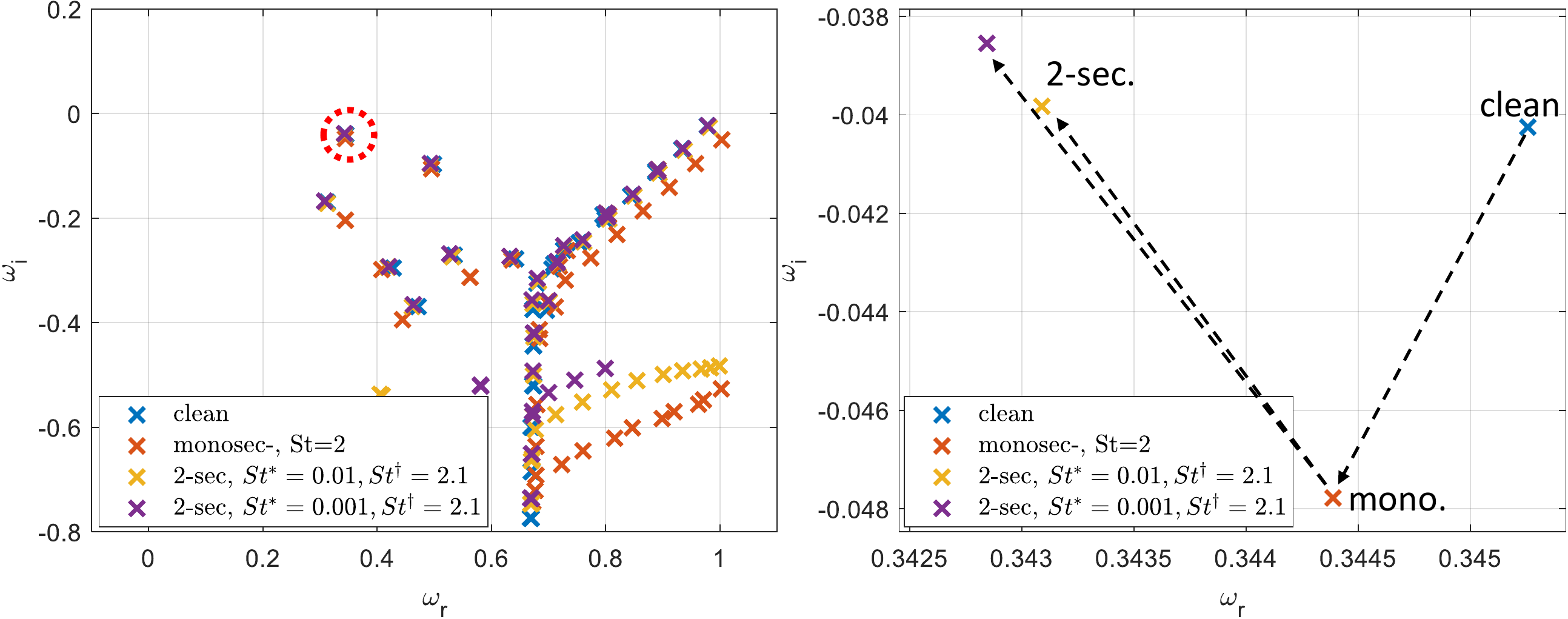}}
	\caption{Stability Spectra of $Re=1000$ clean flow , mono-sectional particle-laden flow with $St=2$, and bisectional particle-laden flow with combinations of: $St^\ast=0.01, St^\dagger=2.1$; $St^\ast=0.001, St^\dagger=2.1$. The four cases display similar shapes with a small shift to the lower-right for the mono-sectional $St=2$ case.  (b) a zoom-in of the most unstable mode of the $A$ family (dashed red circle of the left branch); for this mode, the addition of particles, mono-sectional or two-sectional, lowers the frequency. Mono-sectional particles stabilize the flow whereas two-sectional particles slightly destabilize it.}
	\label{fig:yshape1Kst2}
\end{figure}

\begin{figure}
	\centerline{\includegraphics[width=1\textwidth]{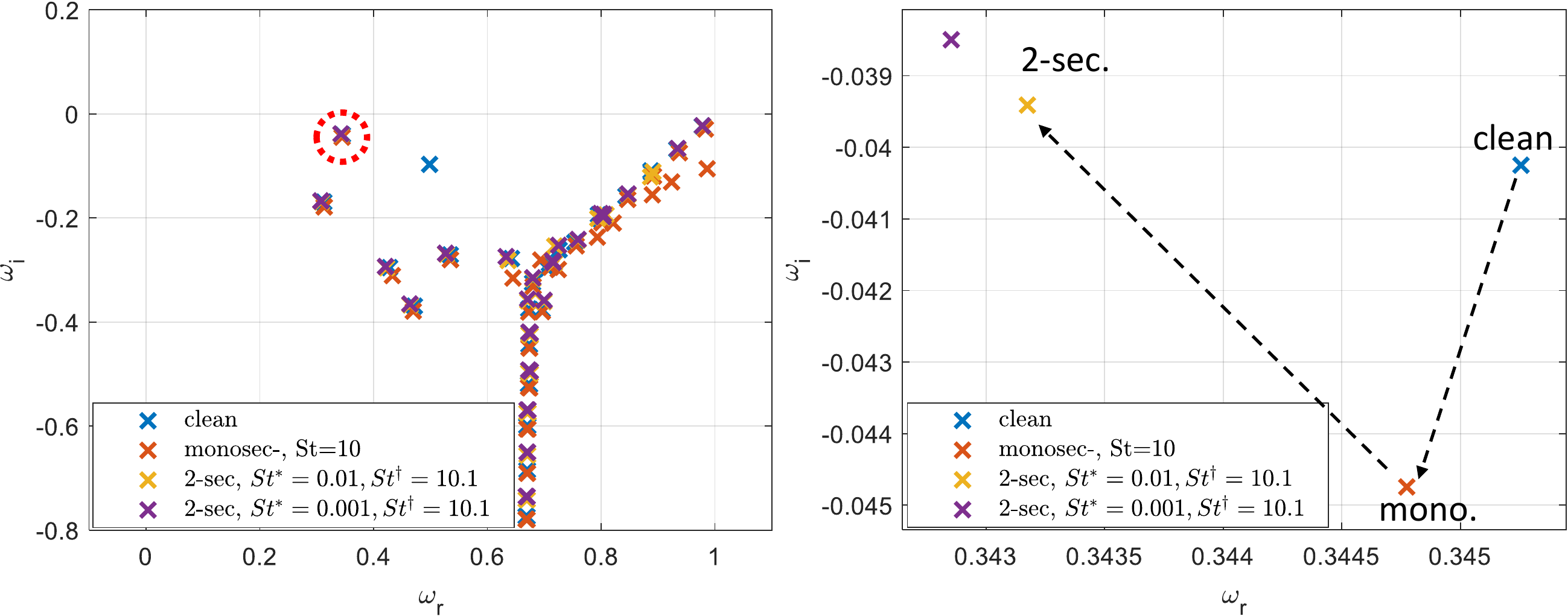}}
	\caption{Stability Spectra of $Re=1000$ clean flow , mono-sectional particle-laden flow with $St=10$, and bisectional particle-laden flow with combinations of: $St^\ast=0.01, St^\dagger=10.1$; $St^\ast=0.001, St^\dagger=10.1$. The four cases display similar shapes with a small shift to the lower-right for the mono-sectional $St=10$ case.  (b) a zoom-in of the most unstable mode of the $A$ family (dashed red circle of the left branch); for this mode, the addition of particles, mono-sectional or two-sectional, lowers the frequency. Mono-sectional particles stabilize the flow whereas two-sectional particles slightly destabilize it.}
	\label{fig:yshape1Kst10}
\end{figure}

\begin{figure}
 	\centerline{\includegraphics[width=1\textwidth]{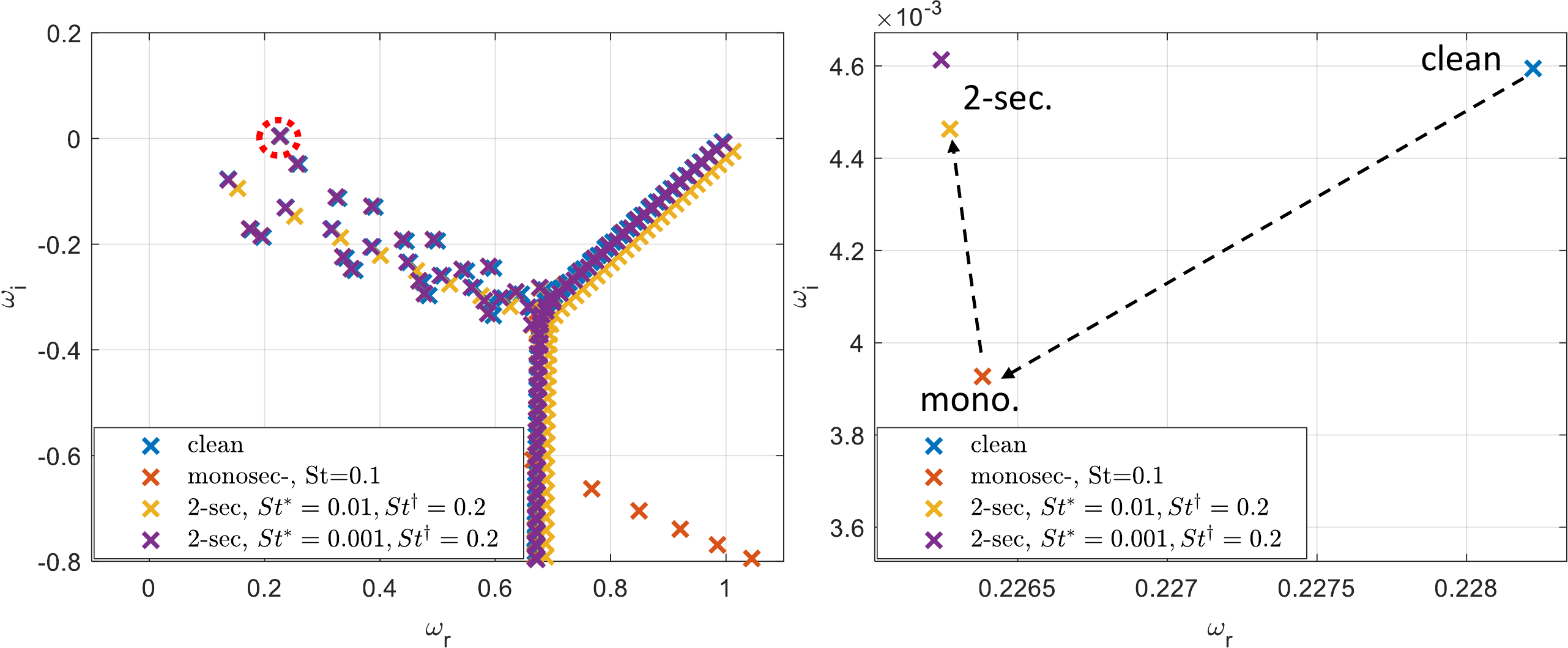}}
 	\caption{Stability Spectra of $Re=12500$ clean flow , mono-sectional particle-laden flow with $St=0.1$, and bisectional particle-laden flow with combinations of: $St^\ast=0.01, St^\dagger=0.2$; $St^\ast=0.001, St^\dagger=0.2$. The four cases are closely aligned with a small shift to the lower-right for the bisectional $St^\ast=0.01, St^\dagger=0.2$ case.  (b) a zoom-in of the most unstable mode of the $A$ family (dashed red circle of the left branch); for this mode, the addition of particles, mono-sectional or bisectional, lowers the frequency. Mono-sectional particles have a stronger stabilizing effect than the bisectional cases.}
 	\label{fig:yshape12Kst0.1}
 \end{figure}

\begin{figure}
 	\centerline{\includegraphics[width=1\textwidth]{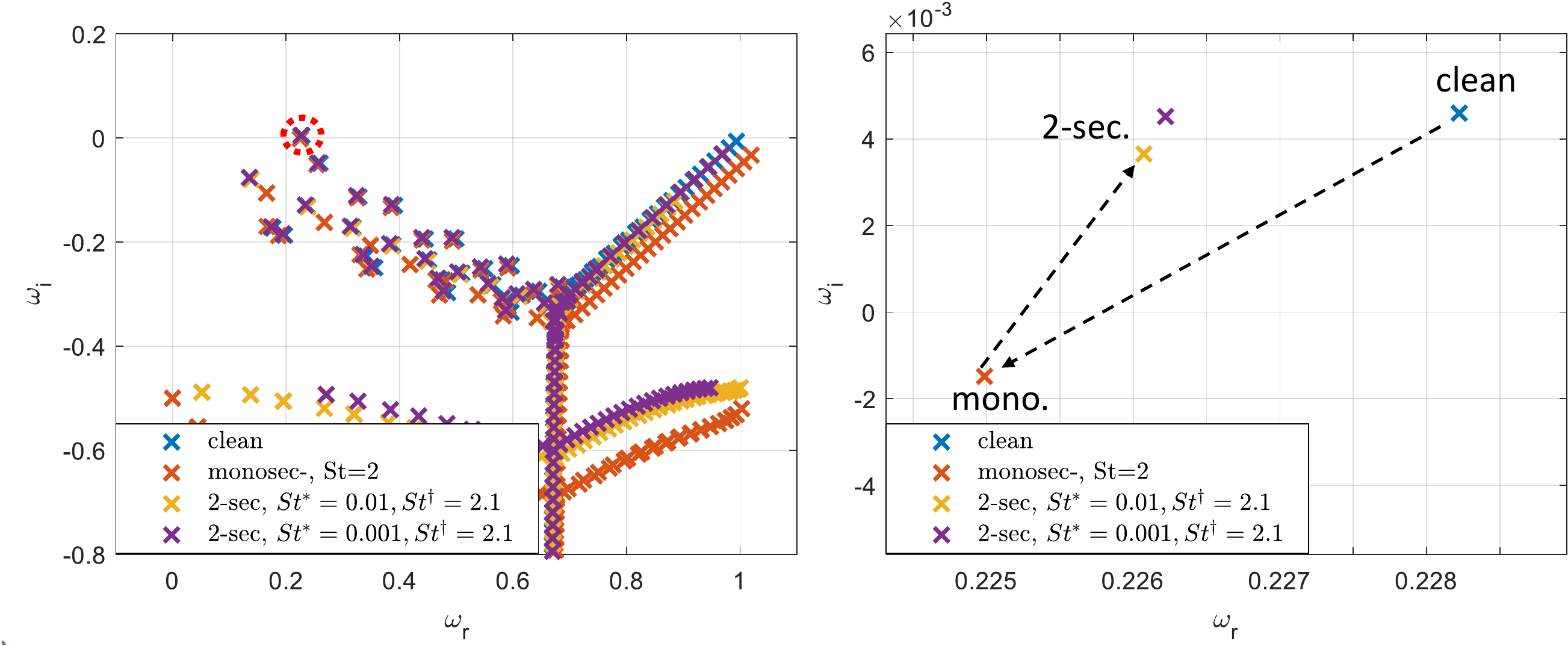}}
 	\caption{Stability Spectra of $Re=12500$ clean flow , mono-sectional particle-laden flow with $St=2$, and bisectional particle-laden flow with combinations of: $St^\ast=0.01, St^\dagger=2.1$; $St^\ast=0.001, St^\dagger=2.1$. The four cases are closely aligned with a small shift to the lower-right for the mono-sectional $St=2$ case. (b) a zoom-in of the most unstable mode of the $A$ family (dashed red circle of the left branch); for this mode,  the addition of particles, mono-sectional or bisectional, lowers frequency. Mono-sectional particles have a stronger stabilizing effect than bisectional cases.}
 	\label{fig:yshape12Kst2}
 \end{figure}

\begin{figure}
 	\centerline{\includegraphics[width=1\textwidth]{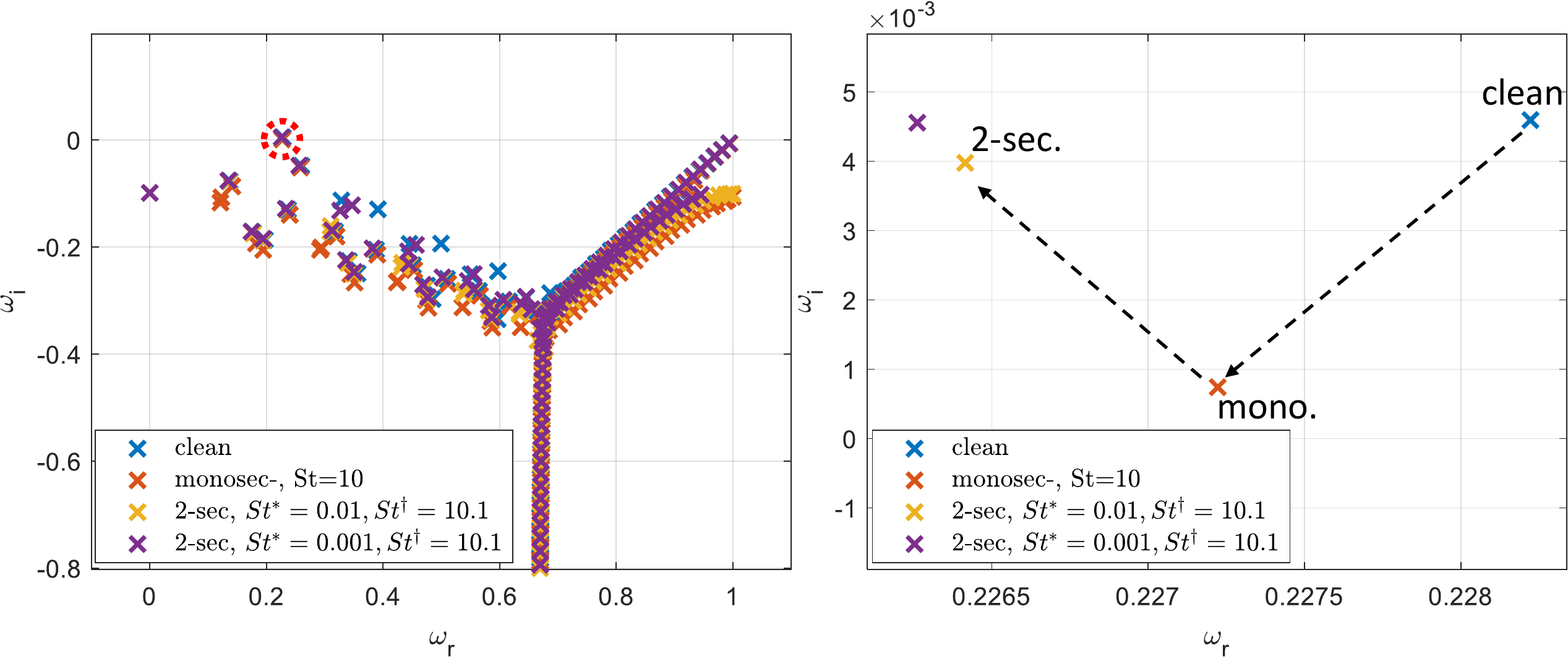}}
 	\caption{Stability Spectra of $Re=12500$ clean flow , mono-sectional particle-laden flow with $St=10$, and bisectional particle-laden flow with combinations of: $St^\ast=0.01, St^\dagger=10.1$; $St^\ast=0.001, St^\dagger=10.1$. The four cases are closely aligned. (b) a zoom-in of the most unstable mode of the $A$ family (dashed red circle of the left branch); for this mode,  the addition of particles, mono-sectional or bisectional, lowers frequency as well as growth rate. Mono-sectional particles have a stronger stabilizing effect than bisectional cases.  }
 	\label{fig:yshape12Kst10}
 \end{figure}

We examine the effects of both monodisperse and polydisperse particles on the stability compared to their carrier-flow clean state. Stability spectra of clean, monodisperse and polydisperse particle-laden flows at $Re=1000$ and $Re=12500$ are shown in figures \ref{fig:yshape1Kst0.1}-\ref{fig:yshape1Kst10} and figures \ref{fig:yshape12Kst0.1}-\ref{fig:yshape12Kst10}, respectively. Specifically, we examine cases of bi-sectional particles ($N_s=2$), which allows us to properly isolate the effect of adding fine dust, or fine mist in the case of droplets, into a particle-laden flow of relatively larger mono-sectional distribution. Note that generally, any number of sections, $N_s$, may be used using our presented approach. Here, $\omega_{r}$, the real part of eigenvalues, represents the real frequency, whereas $\omega_{i}$, the imaginary part of eigenvalues, is the temporal growth rate. While each case has its distinctive stability spectrum, the general influence of fine particle addition can be observed and summarized as follows.

Overall, the eigenvalue maps of all cases - the clean, mono - and bisectional particle-laden flows - look similar. However, a closer look at the most unstable $A$ branch mode shows that: at $Re=1000$, the addition of mono-sectional fine particles ($St\sim O(0.1)$) increases the growth rate as well as frequency. However, the increasing Stokes number, which may be realized by the addition of coarser mono-sectional particles, suppresses the growth rate with respect to the carrier-flow clean state. Notably, particles of bi-sectional distribution increase the growth rate and decrease the frequency for all examined Stokes numbers. On the other hand, observing the right branch of the eigenvalue map, the $P$ family eigenvalues, we see that the addition of different cases of particles results in a shift to higher frequencies and lower growth rates compared to the clean flow.

Increasing the Reynolds number to $Re=12500$, all cases of particle additions result in lowering frequency and most have stabilizing effects (figures \ref{fig:yshape12Kst0.1}-\ref{fig:yshape12Kst10}). The most notable stabilization occurs in the monodisperse case for $St=2$. Notably, for the polydisperse cases of $Re=12500$, $St^\ast=0.001$ leads to a slightly less stable flow than $St^\ast=0.01$, and its growth rate is similar to that of the clean flow thereof.

Next, the most unstable modes on the $A$ family of the stability spectra are extracted and discussed. The growth rate characteristics for monodisperse particle-laden flows of $Re=12500$ and $Re=1000$  are shown in figures ~\ref{fig:comparisona} and~\ref{fig:lowRe} and, respectively; the monodisperse particle-laden flow becomes most unstable at relatively smaller Stokes number. Its stability increases as the Stokes number increases up to approximately $4$, at which the flow is most stabilized compared to its clean state. Further increasing the Stokes number, the particle-laden flow grows unstable again and slowly approaches its clean state maximal growth rate.

The effect of adding dusty particles into a flow already laden with larger size particles is now examined by analysing the two-section case by setting $N_s=2$ and $j=1,2$ in equations (\ref{eq:gasmomori}) through (\ref{eq:parmomori}). 
To decide the mass fraction of particles in two distinct size sections, $Q_{0}^\ast$ and $Q_{0}^\dagger$, we refer to the SMD defined previously to adjust the desired mean Stokes number of the combined bi-sectional particle distribution.

\begin{figure}
	\centerline{\includegraphics[width=1\textwidth]{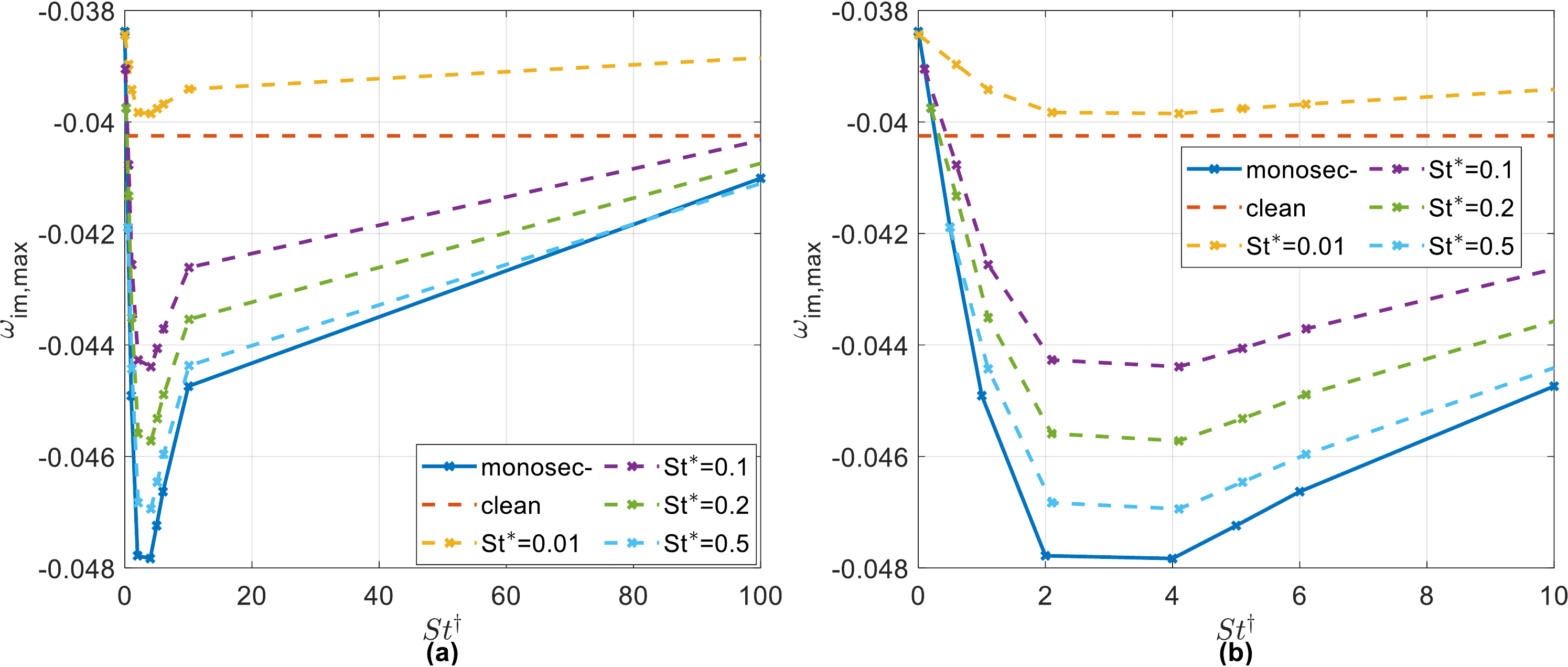}}
	\caption{Particle laden channel flow growth rates for $Re=1000$. The most unstable $A$ branch mode is plotted as a function of particle Stokes number for different cases. The horizontal dashed red line represents the result of the clean carrier flow - unaffected by $St$. The solid blue line represents the results of monodisperse particle addition: particles with $St\geq0.5$ stabilize and particles with $St\leq0.2$ destabilize the flow with respect to its clean state. The dashed lines are constant $St^\ast$ lines, with which selected $St^\dagger$ particles are paired to yield equivalent Stokes numbers as those examined in mono-size particles.. (b) a zoom-in of (a) at the low $St$ regime.}
	\label{fig:monopolyRe1k}
\end{figure}

\begin{figure}
	\centerline{\includegraphics[width=1\textwidth]{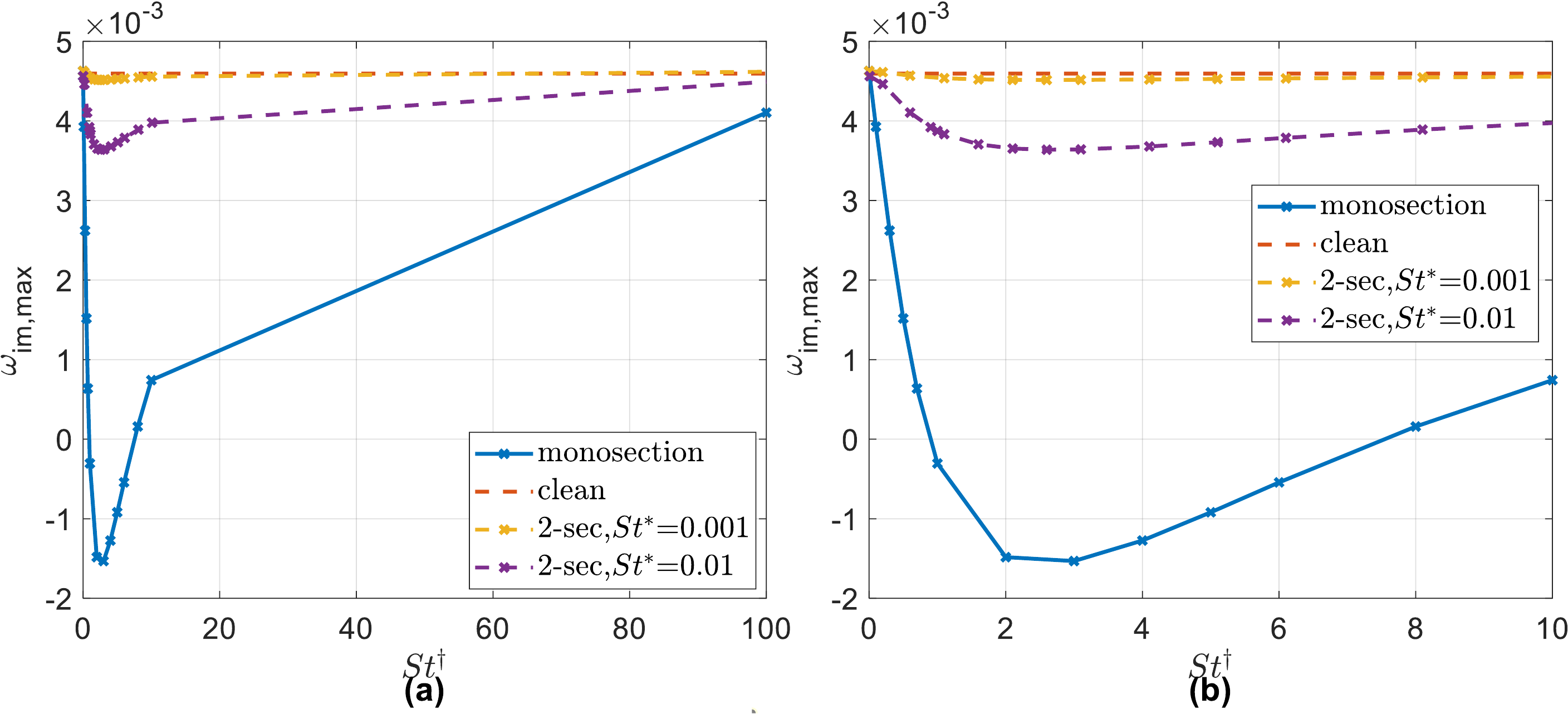}}
	\caption{Particle laden channel flow growth rates at $Re=12500$. The most unstable $A$ branch mode is plotted as a function of particle Stokes number for different cases. The horizontal dashed line represents result of the clean carrier flow. The solid blue line represents results of mono-sized particle addition. The dashed lines are constant $St^\ast$ lines, with which a selected $St^\dagger$ particles are paired to yield equivalent Stokes numbers as those examined in mono-size particles. Plot (b) presents a zoom-in of plot (a) at the low $St$ regime.}
	\label{fig:monopolyRe12k}
\end{figure}

Figure \ref{fig:monopolyRe1k}a shows that for both mono- and bi-sectional cases, the maximal growth rates of the $A$ family eigenvalues converge as the Stokes number increases to $100$. Here, however, we focus on the low Stokes region, which is presented in \ref{fig:monopolyRe1k}b. The combinations of two-section particles with different Stokes numbers, typically $St^\ast \leq 0.5$ and $St^\dagger >0.5$ as appeared in equation \ref{eq:defSMD}, yield different instability growth rates compared to those of the monodisperse cases with the same resulting $St^{SMD}$. The same analysis for $Re=12500$ is presented in figure \ref{fig:monopolyRe12k}, where growth rates increase across the neutral $\omega_{im,max}=0$ line, and thus the flow perturbations may grow exponentially.
This demonstrates how sensitive the addition of fine dust particles to a monodisperse flow is in terms of linear stability.


An \textit{equivalent Stokes number} is now defined such that a given bi-sectionally dispersed system will have the same temporal growth rate as an equivalent monodisperse case with $St^{eq}$.

\begin{figure}
	\centerline{\includegraphics[width=0.6\textwidth]{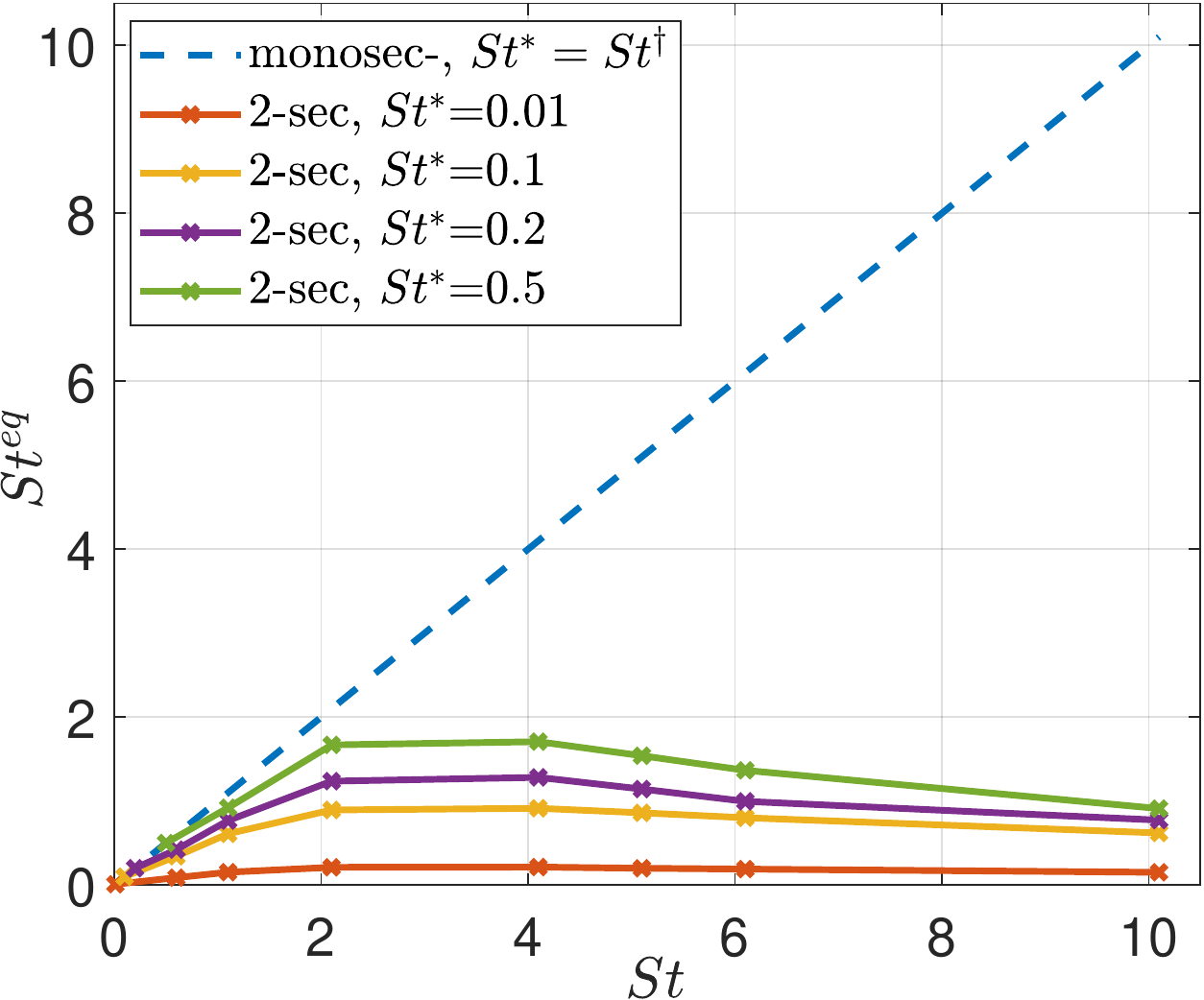}}
	\caption{$St^{eq}$ of several cases of the polydisperse $Re=1000$ channel flows plotted in figure \ref{fig:monopolyRe1k}. The dashed line represents the mono-sectional cases where the equivalent Stokes number equals the Stokes number of the mono-section particles. Each solid line represents a certain value of $St^\ast$, for which different $x$ symbols represent different values of $St^\dagger$. The examined values of $St^\ast$ are $0.01, 0.1, 0.2$ and $0.5$, while the values of $St^\dagger$ are chosen to be $0.6, 1.1, 2.1,4.1,5.1,6.1$ and $10.1$ resulting in $St^{SMD}=0.5, 1, 2, 4, 5, 6$ and $10$. }
	\label{fig:steqRe1k}
\end{figure}

Details of the cases plotted in figure \ref{fig:monopolyRe1k} and their corresponding equivalent Stokes numbers are summarized in the table \ref{table:steq}. In the table, mass fractions of bi-sectional particle distribution are listed under their corresponding section number $j$, with each section characterised by a specific Stokes number, along with the equivalent Stokes number of each case. Note that two such equivalent Stokes numbers may be realized for each monodisperse curve, which appears in figure \ref{fig:monopolyRe1k} - one in the lower $St$ region on the left side and the other one in the higher $St$ region on the right side. 
Here we choose to focus only on the lower $St$ region and thus choose the $St^{eq}$ on the left side of the curve. Figure \ref{fig:steqRe1k} summarizes the equivalent Stokes numbers in polydisperse cases with values of $St^\ast=0.01, 0.1, 0.2, 0.5$ and $St^\dagger=0.6, 1.1, 2.1,4.1,5.1,6.1,10.1$, which together yield the mean Stokes number of $St^{SMD}=0.5, 1, 2, 4, 5, 6, 10$, respectively. The addition of a small amount of finer particles lowers the stabilizing effect of larger particles, resembling a mono-sectional case of a lower Stokes number. This effect is strengthened as $St^\dagger$ increases, indicated by the solid curves deviating farther from the dashed straight line as shown in figure \ref{fig:steqRe1k}.

\begin{table}
\centering
	\begin{tabular}{cccccccccc}
		\hline
		& j=1      & j=2 & j=3      & j=4 & j=5     & j=6 & j=7      & $St^{SMD}$ & $St^{eq}$ \\
		St     & 0.01     & 0.1 & 0.2      & 2   & 2.1     & 10  & 10.1     &            &           \\
		case 1 & 0        & 1   & 0        & 0   & 0       & 0   & 0        & 0.1        & 0.1       \\
		case 2 & 0.119296 & 0   & 0.880704 & 0   & 0       & 0   & 0        & 0.1        & 0.0248    \\
		case 3 & 0        & 0   & 0        & 1   & 0       & 0   & 0        & 2          & 2         \\
		case 4 & 0.00183  & 0   & 0        & 0   & 0.99817 & 0   & 0        & 2          & 0.2112    \\
		case 5 & 0        & 0   & 0        & 0   & 0       & 1   & 0        & 10         & 10        \\
		case 6 & 0.000162 & 0   & 0        & 0   & 0       & 0   & 0.999838 & 10         & 0.1515    \\ \hline
	\end{tabular}
\caption{Example cases of monodisperse and polydisperse bisectional particle-laden flows with $St^\ast=0.01$. The mean polydisperse Stokes numbers $St^{SMD}$ are listed next to the equivalent Stokes number $St^{eq}$ (also shown in figure  \ref{fig:monopolyRe1k}).}
\label{table:steq}
\end{table}

\subsection{Spatial wavenumber effect} 
\begin{figure}
	\centerline{\includegraphics[width=1\textwidth]{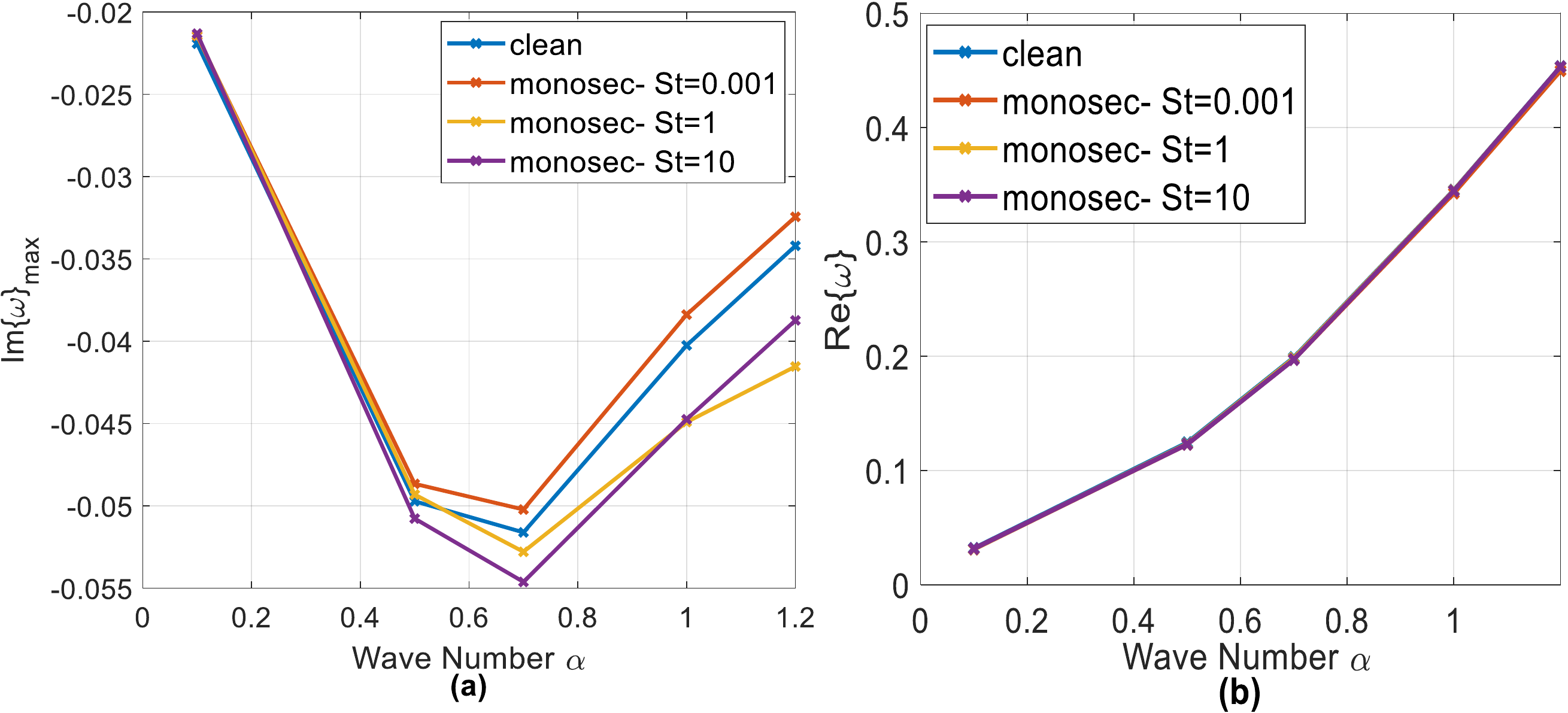}}
	\caption{The most unstable $A$ family branch eigenvalues as a function of streamwise wavenumber $\alpha$. (a) and (b) subplots show the variation of the maximal imaginary part (perturbation growth rate), $\operatorname{Im}\{\omega\}_{max}$, and the corresponding real part (perturbation frequency), $\operatorname{Re}\{\omega\}$, respectively. Four channel flow cases are plotted for $Re=1000$: clean state, mono-section with $St=0.001$, $St=1$ and $St=10$.  }
	\label{fig:alpom}
\end{figure}

The effects of varying the streamwise wavenumber $\alpha$ on the most unstable $A$ family branch eigenvalues are investigated, as shown in figure \ref{fig:alpom}. Four cases of $Re=1000$ channel flows are calculated and compared: clean flow, and monodisperse flows laden with $St=0.001$, $St=1$ and $St=10$ particles. The maximal growth rates differ slightly with increasing $\alpha$, and the $St=0.001$ monodisperse case is consistently the most unstable, while monodisperse particles of $St=1$ and $St=10$ tend to be more stable than the clean flow for $\alpha>0.6$. The corresponding real frequencies monotonically increase with $\alpha$, and seem to be unaffected by particle addition throughout the investigated range of wavenumbers.

\subsection{Perturbations Eigenmodes}
Eigenfunctions of the most unstable $A$ family eigenvalues in clean gas flow, mono-section and 2-section particle-laden flows at $\alpha=1$, $Re=1000$ as well as $Re=12500$ are studied and compared. Eigenmodes of the carrier flow perturbations $u^\prime, v^\prime, p^\prime$ and eigenmodes of sectional particles $u_j^\prime, v_j^\prime, q_j^\prime (j=1,..., N_s)$ can be calculated from the system of eigenvalue problem presented in section \ref{section:normalmodes}. Previously, we showed in figure \ref{fig:monopolyRe1k} that very fine particles destabilize the mono-section particle-laden flows compared to its clean state, while particles of $St=2-4$ have the most pronounced stabilizing effect; and as Stokes number further increases, growth rates increase. Based on these observations, eigenfunctions in cases with particles of $St=0.1$, $St=2$ and $St=10$ are selected here for comparison to understand their stability characteristics.

\begin{figure}
	\centerline{\includegraphics[width=1\textwidth]{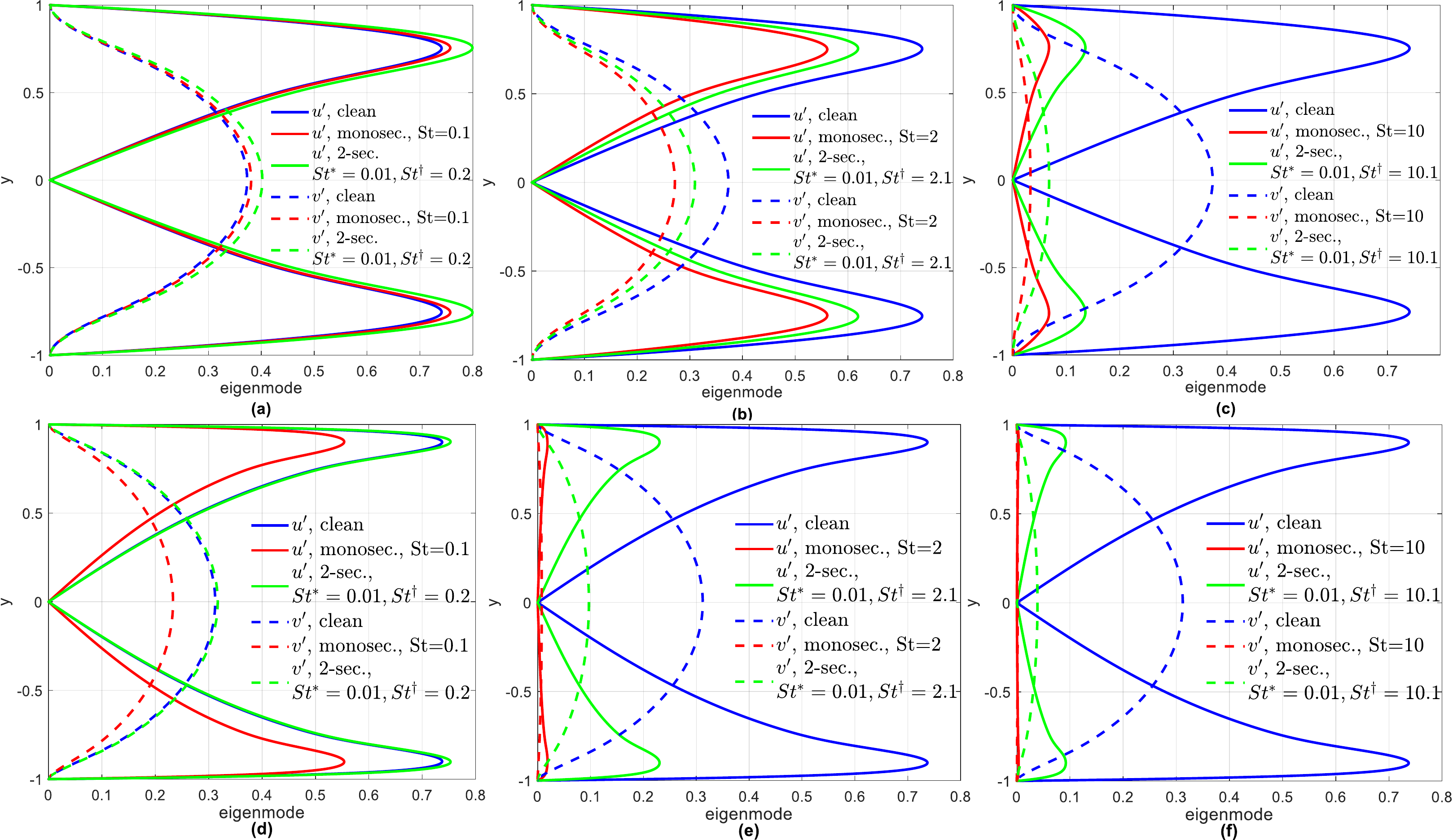}}
	\caption{Carrier-flow velocity eigenfunctions of the most unstable $A$ branch mode for $Re=1000$ $\left(\left(a\right)-\left(c\right)\right)$, $Re=12500$ $\left(\left(d\right)-\left(f\right)\right)$ and loading of $Q_{0,tot}=0.05$ for clean flows, mono-sectional and bi-sectional particle laden flows with SMD Stokes numbers of $St^{SMD}=0.1$ (a) and (d),  $St^{SMD}=2$ (b) and (e), and  $St^{SMD}=10$ (c) and (f),  respectively. All bi-sectional distributions comprise of finer particles of  $St^\ast=0.01$.}.
	\label{fig:efuv}
\end{figure}

Figure \ref{fig:efuv}(a) shows that the magnitudes of perturbations of carrier-flow streamwise velocity $|u^\prime|$ and vertical velocity $|v^\prime|$ are similar for the clean flow, and the $St^{SMD}=0.1$ particle-laden flow cases, reaching maximal amplitudes at the same distance from the wall. The two-section particle-laden flow has the largest magnitudes of perturbations, while the clean flow has the smallest. In figure \ref{fig:efuv}(b), mono-sectional and polydisperse particles with $St^{SMD}=2$ effectively reduce the carrier flow perturbation magnitudes. Furthermore, mono-sectional and polydisperse particles with $St^{SMD}=10$ significantly reduce carrier-flow perturbation from $0.74$ to lower than $0.15$ for $|u^\prime|$, and from $0.37$ to lower than $0.07$ for $|v^\prime|$. The maximal perturbation occurs at approximately the same location for all cases. 

The same three cases of $Re=12500$ are shown in figure \ref{fig:efuv} $(d)-(f)$, in which the locations of the maximum perturbations are shifted towards the channel walls compared to the case of $Re=1000$. In general, for  $Re=12500$, perturbation magnitudes are smaller than those of $Re=1000$, which is explained by lower absolute values of the corresponding growth rates. The carrier-flow velocity perturbations are mostly decayed by the addition of $St=10$ monodisperse particles. On the contrary, the addition of finer particles will slightly increase the carrier-flow perturbations as indicated in figure~\ref{fig:efuv}$(a$ and $d)$. \\

\begin{figure}
	\centerline{\includegraphics[width=1\textwidth]{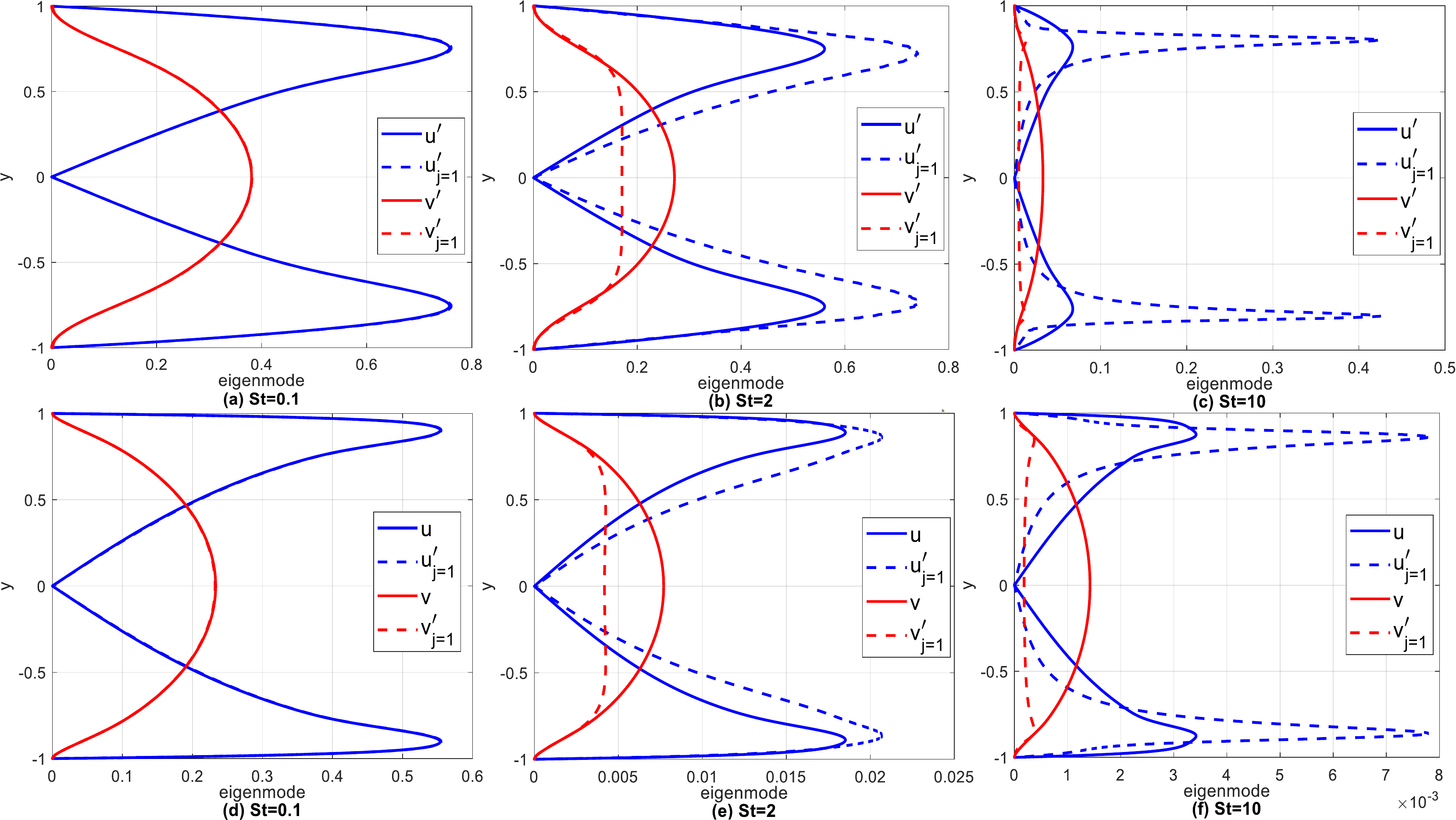}}
	\caption{Carrier-flow velocity and particle velocity eigenfunctions of the most unstable $A$ branch mode presented for mono-sectional particles and $Q_{0,tot}=0.05$. (a-c) Stokes numbers $St=0.1$,  $St=2$ and  $St=10$, respectively and $Re=1000$; (d-f) same as (a-c) Stokes numbers $St=0.1$,  $St=2$ and  $St=10$, respectively and $Re=12500$.}
	\label{fig:ef_upvp_mono}
\end{figure}

Eigenmodes of particle velocity perturbations in monodisperse channel flows are given in figure \ref{fig:ef_upvp_mono}, where $u_1^\prime$ and $v_1^\prime$ are the monodisperse particle velocity perturbations, and $u^\prime$ and $v^\prime$ are those of the carrier flow.
For $St=0.1$, the difference between particles and carrier-flow perturbed velocities is hardly noticeable because the particles are small enough to follow the flow stream-lines. However, for larger Stokes numbers, the particle streamwise velocity perturbation $u_1^\prime$ is larger than that of the carrier flow, and the particle vertical velocity perturbation $v_1^\prime$ is smaller than that of the carrier flow, for both $Re=1000$ and $Re=12500$ cases. Comparing figures~\ref{fig:ef_upvp_mono}(a-c), the monodisperse particle Stokes number increases, the difference between the two phases' velocity perturbation grows, which represents the effect of the linear drag between carrier flow and particles according to equation \ref{eq:pert_u} to \ref{eq:pert_vj}.
With increasing $St$ of the monodisperse case,  velocity perturbation magnitudes decrease for both the carrier flow and particulate matter. Moreover, the modal shape of $u_1^\prime$ becomes narrower.
For the case of a high Reynolds number ($Re=12500$), the same shapes are observed with maximum perturbation locations shifted towards channel walls compared to the $Re=1000$ case. The magnitudes of perturbations become extremely small with the addition of particles of $St=10$ compared to those of $St=0.1$.\\

\begin{figure}
	\centerline{\includegraphics[width=1\textwidth]{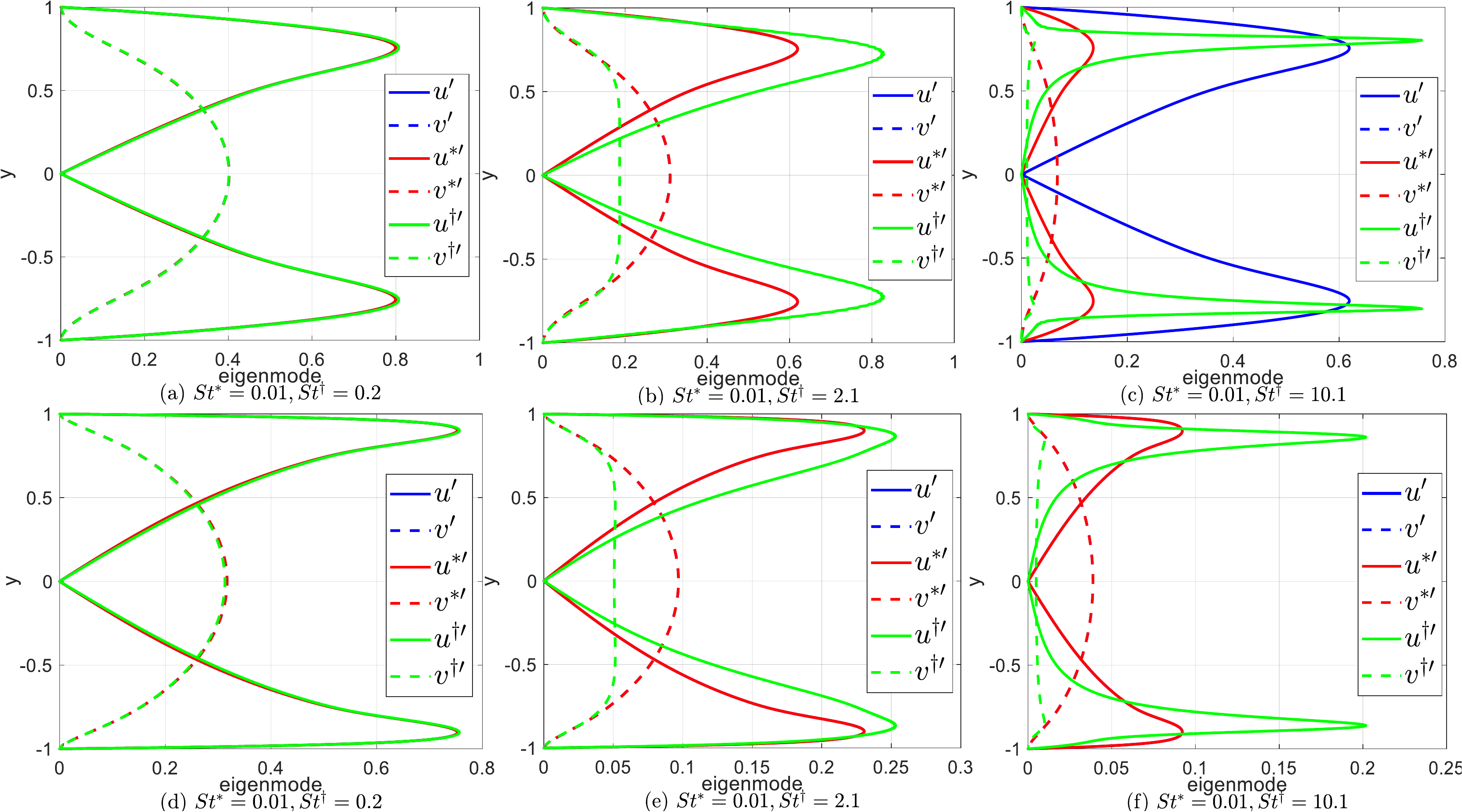}}
	\caption{Eigenmodes of bi-sectional particle laden flows with combinations of $St^\ast=0.01$ and different Stokes numbers $St^\dagger$, and $Re=1000$ ((a)-(c)), $Re=12500$ ((d)-(f)) respectively. Carrier-flow velocity perturbations $u^\prime, v^\prime$ and two section particles velocity perturbations $u^{\ast \prime}, v^{\ast \prime}, u^{\dagger \prime}$ and $v^{\dagger \prime}$ are plotted for each case.}
	\label{fig:ef-twoperts}
\end{figure}

Figure \ref{fig:ef-twoperts} shows the eigenfunctions of a bi-section particle-laden flows at different Stokes numbers, $Re=1000$ ((a)-(c)), and $Re=12500$ ((d)-(f)), respectively. Eigenfunctions for the carrier-flow velocity perturbations $u^\prime, v^\prime$ and particles of sections $St^\ast$ and $St^\dagger$ velocity perturbations, $u^{\ast\prime}, v^{\ast\prime}$ and $u{\dagger\prime}, v{\dagger\prime}$, are compared. In all cases except for figure~\ref{fig:ef-twoperts}c, particles of $St^\ast=0.01$ have the same velocity perturbation magnitudes as those of the carrier flow. 
Especially when the coarser particles are also relatively small, $St^\dagger=0.2$ as in plots (a) and (d), the two phases have the same velocity perturbation magnitudes, and relative motions are hardly noticeable, and thus, less drag-induced dissipation of energy is expected \citep{klinkenberg2011modal}. 

As $St^\dagger$ increases to $St^\dagger=2$ and $St^\dagger=10$, coarser particles play the role of weakening the carrier-flow velocity perturbations $u^\prime$ and $v^\prime$, similar to the cases demonstrated in figure \ref{fig:ef_upvp_mono} for the mono-sectional cases. We can also observe that as $St^\dagger$ increases, the magnitudes of $u^{\ast \prime}$ and $v^{\ast \prime}$ (the velocity perturbations of the fine $St^\ast$ particles) significantly decreases, and that the shape of $u^{\dagger \prime}$ (the velocity perturbation of coarser $St^\dagger$ particles) becomes narrower.

\begin{figure}
	\centerline{\includegraphics[width=1\textwidth]{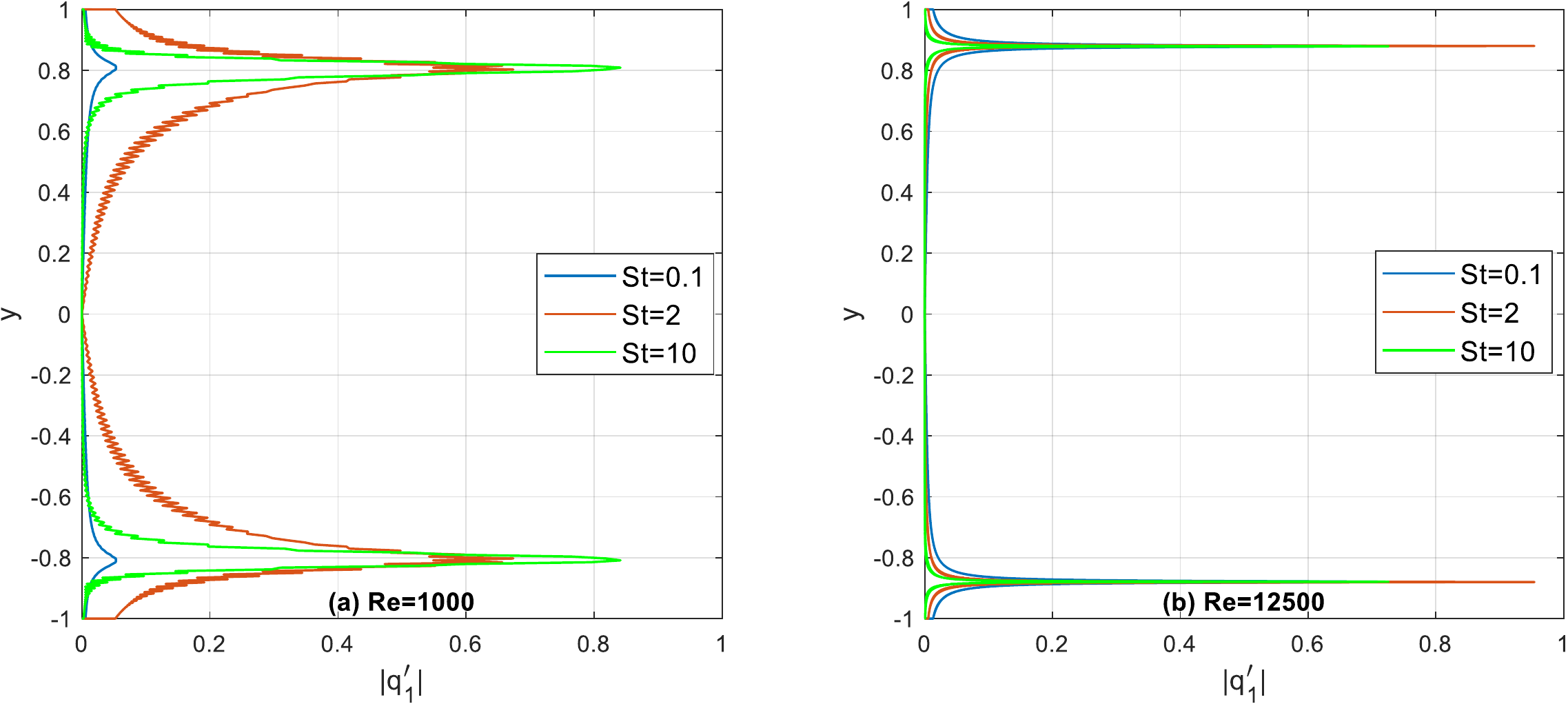}}
	\caption{Perturbation magnitude profile of monodisperse particle mass fraction $q_1^\prime$ for $Re=1000$ (a) and for $Re=12500$ (b). Three sizes of monodisperse particles are investigated: $St=0.1$, $St=2$ and $St=10$.}.
	\label{fig:ef-monoq}
\end{figure}

Perturbations in particle loading $|q^\prime_1|$ in three monodisperse cases are given in figure \ref{fig:ef-monoq}. 
The largest magnitude of $q^\prime_1$ is obtained at approximately the same location as $u^\prime_1$ in figure \ref{fig:ef_upvp_mono}, increasing with the increase of Stokes number of the mono-sectional particles. The conspicuous fluctuations of the $Re=1000, St=2$ monodisperse case may be related to the imposition of boundary condition $q^\prime_1=0$ at the walls. Similar fluctuations have also been observed in the carrier-flow pressure perturbation $p^\prime$.

\section{Concluding Remarks}
A new generalized mathematical framework is developed for the hydrodynamic stability of polydisperse particle-laden shear flows, whereby inter-phase and inter-particle interactions may be realized.
The flow and dispersed particulate matter are described using this approach by combining a linear modal stability analysis and a discrete Eulerian sectional formulation. Separate transport equations are formulated for each size section of the dispersed phase, where interphase and inter-particle mass and momentum transfer are
modelled as source terms in the governing carrier flow and particulate phase equations. 

In this study, a linear modal stability framework was developed by linearizing the coupled flow and particle equations in a Cartesian frame of reference to study polydisperse particle-laden channel flows, where a generalized eigenvalue problem incorporating the carrier flow equations and terms for each size-section of the dispersed phase is derived.

The stability characteristics of clean and monodisperse particle-laden channel flow using our approach are validated with those obtained using the Orr-Sommerfeld equation for clean flows and with the results reported by~\cite{klinkenberg2011modal}.
We find that the monodisperse channel flow stability is the most stable around Stokes number of unity, whereas lower and higher Stokes numbers exhibit higher growth rates that may exceed the clean flow growth rates. 

By analysing the eigenvalue maps of monodisperse particles, we find that the addition of monodisperse particles of $St= \mathcal{O}(0.1)$ shifts the most unstable $A$ branch eigenvalue toward higher growth rates for $Re=1000$. However, adding the same Stokes number particles with $Re=12500$ tends to lower the growth rates of the most unstable $A$ branch eigenvalue, indicating a more stable flow; this reveals the complex interplay between flow and monodisperse particle time-scales and their effect on the stability characteristics.

By adding fine particles into the monodisperse particle-laden flow, we study for the first time the instability of a polydisperse flow configuration. 
We find that instability may drastically increase compared to the monodisperse case, and growth rates may even surpass those of the clean flow.
Based on the eigenvalue growth rates, an equivalent Stokes number is devised, for which the growth rates of the polydisperse system are the same as that of an equivalent mono-sectional particle distribution. 
This allows a comprehensive comparison between monodisperse and polydisperse flow systems in terms of stability.
The lower the fine particle addition Stokes number $St^\ast$ is, the more significant the destabilizing effect is, thus deviating the equivalent Stokes number, $St^{eq}$, away from the value of coarser particles $St^\dagger$. 

Clean flow and monodisperse flows of different Stokes numbers present a similar stability trend as a function of axial wavenumber $\alpha$. Eigenfunctions of perturbation profiles show that Monodisperse and bi-sectional distributed particles with $St^{SMD}=0.1$ slightly increase the gaseous flow velocity perturbations magnitudes, while mono-sectional and bi-sectional particles with $St^{SMD}=10$ significantly reduce the perturbation magnitudes. Perturbations of monodisperse particle velocities are different from those of the carrier gas flow, especially for $St=10$, the particulate matter experiences a significantly larger magnitude of axial velocity perturbation than that of the carrier flow.

Although the present study focuses on a non-interacting multiphase flows, our new mathematical approach has extended the hydrodynamic stability framework by including a mechanism for inter-phase and inter-sectional interactions, which is generally described here in sections~\ref{sec:mathmodelling}. 
The utility of our approach in rationalizing more complex polydisperse systems, including evaporation, condensation, and droplet coalescence, is currently being examined. 


\section*{Acknowledgement}
This research was supported by the ISRAEL SCIENCE FOUNDATION (grant No. 1762/20).

\section*{Declaration of Interests} 
The authors report no conflict of interest.


\newpage
\section{Appendix}
\label{section:app}
The generalized eigenvalue problem in this study is defined by $\mathbf{A X}=\omega \mathbf{B X}$ with $\omega$ being the eigenfrequencies and $\mathbf{X}$ being the eigenvector column in equation \ref{eq:eigfunc}. The following will describe the two matrices in detail. 
\[
A=\begin{bmatrix}
\centering
A_{0,0}& A_{0,1} & A_{0,2} & ... & A_{0,j}& ... &A_{0,N_s} \\
A_{1,0}& A_{1,1} & A_{1,2} & [0] & ... & ... &[0]\\ 
A_{2,0}& [0] & A_{2,2} & A_{2,3} & [0] & ... &[0]\\
... & ... & ... & ... & ... & ... &... \\
A_{j,0} & [0] & [0] & A_{j,j} & A_{j,j+1} & [0] &[0]\\
... & ... & ... & ... & ... & ... &...\\
A_{N_s,0} & [0] & ... & ... & ... & [0] &A_{N_s,N_s}\\
\end{bmatrix}
\]
The subscripts in small matrices $[A_{a,b}]$ represent the contribution of section $b$ to section $a$'s equations. Number "0" denotes the carrier-flow, while numbers "1"-"$N_s$"
denote the specific particulate matter section number, i.e. $j=1,2,...,N_s$. We can further divide the small matrices $[A_{a,b}]$ into 9 even smaller matrices $[A_{axby}]$ with $x=1,2,3$ and $y=1,2,3$. The index $x$ denotes the three equations (one for mass conservation and two for momentum conservation) of section $a$ and the index $y$ denotes the three eigenfunctions of section $b$, respectively. We construct matrix $[A_{a,b}]$ in such a way as
\[
A_{a,b}=\begin{bmatrix}
\centering
A_{a1b1}& A_{a1b2} &A_{a1b3}\\
A_{a2b1}& A_{a2b2} &A_{a2b3}\\
A_{a3b1}& A_{a3b2} &A_{a3b3} \\
\end{bmatrix}
\]
In particular, the three eigenfunctions for carrier-flow, section "0", are $\{F(y), G(y), P(y)\}$, and those for sectional particulate matter $j=1,2,...,N_s$ are $\{M_j(y), F_j(y), G_j(y)\}$. For example, matrix $[A_{01j2}]$ denotes the contribution of the eigenfunction $F_j$ from section $j$ to the flow mass conservation equation. The smaller non-zero matrices $[A_{axby}]$ can be extracted from equations \ref{eq:efgasmass_ch}-\ref{eq:efparv} and are presented in details as follows.

The \textit{\textbf{D}} matrix is the differentiation matrix generated by the Chebyshev collocation method provided by \cite{trefethen2000spectral}. Matrices $\mathit{\mathbf{U}}$, $\mathit{\mathbf{U_j}}$, $\mathit{\mathbf{V}}$, and $\mathit{\mathbf{V_j}}$ are diagonal matrices of base-flow velocity at each collocation point. $\mathit{\mathbf{I}}$ is the identity matrix.
\begin{equation*}
    \left[A_{0101}\right]=i\alpha \textit{\textbf{I}} \hspace{0.2in} \left[A_{0102}\right]=i \textit{\textbf{D}} \hspace{0.2in}\left[A_{0103}\right]=\textit{\textbf{0}}
\end{equation*}
\begin{equation*}
    \left[A_{0201}\right]=i\alpha \textit{\textbf{U}}-\frac{1}{Re}\left(\textit{\textbf{D}}^2-\alpha^2\right)+\sum^{N_s}_{j=1}\frac{Q_{0,j}}{St_j}\cdot \textit{\textbf{I}}
    \hspace{0.2in} 
    \left[A_{0202}\right]=i \frac{\partial \textit{\textbf{U}} }{\partial y} \hspace{0.2in}\left[A_{0203}\right]=i\alpha \textit{\textbf{I}}
\end{equation*}
\begin{equation*}
    \left[A_{0301}\right]=\textit{\textbf{0}} 
    \hspace{0.2in} 
    \left[A_{0302}\right]=-\alpha \textit{\textbf{U}}-\frac{i}{Re}\left(\textit{\textbf{D}}^2-\alpha^2\right)+i\sum^{N_s}_{j=1}\frac{Q_{0,j}}{St_j}\cdot \textit{\textbf{I}} \hspace{0.2in}\left[A_{0303}\right]=\textit{\textbf{D}} 
\end{equation*}
for $j=1, 2,..., N_s$:
\begin{equation*}
    \left[A_{01j1}\right]=-C_j \textit{\textbf{I}}
\end{equation*}
\begin{equation*}
    \left[A_{01(j+1)1}\right]=B_{j,j+1} \textit{\textbf{I}} ~~~ \text{(vanishes when $j=N_s$)}
\end{equation*}
\begin{equation*}
    \left[A_{02j1}\right]=-\frac{1}{St_j}\left(\mathbf{U_j}-\mathbf{U}\right)
    \hspace{0.2in}
    \left[A_{02j2}\right]=-\frac{Q_{0,j}}{St_j}\textit{\textbf{I}}
\end{equation*}
\begin{equation*}
    \left[A_{03j1}\right]=-\frac{1}{St_j}\left(\mathbf{V_j}-\mathbf{V}\right)
    \hspace{0.2in}
    \left[A_{03j3}\right]=-i\frac{Q_{0,j}}{St_j}\textit{\textbf{I}}
\end{equation*}
\begin{equation*}
    \left[A_{j201}\right]=-\frac{Q_{0,j}}{St_j}\textit{\textbf{I}}\hspace{0.2in}
    \left[A_{j302}\right]=-i\frac{Q_{0,j}}{St_j}\textit{\textbf{I}}
\end{equation*}
\begin{equation*}
    \left[A_{j1j1}\right]=\left( \frac{\partial\mathbf{U_j}}{\partial x}+\frac{\partial \mathbf{V_j}}{\partial y}+i\alpha \mathbf{U_j}+\mathbf{V_j} \textbf{\textit{D}}+C_j\textbf{\textit{I}}\right)
\end{equation*}
\begin{equation*}
    \left[A_{j1j2}\right]=\left( i\alpha Q_{0,j}+\frac{\partial Q_{0,j}}{\partial x}\right)\textit{\textbf{I}}
    \hspace{0.2in}
    \left[A_{j1j3}\right]=i Q_{0,j}\textbf{\textit{D}}+i\frac{\partial Q_{0,j}}{\partial y}\textbf{\textit{I}}
\end{equation*}
\begin{equation*}
    \left[A_{j1(j+1)1}\right]=-B_{j,j+1} \textit{\textbf{I}} ~~~ \text{(vanishes when $j=N_s$)}
\end{equation*}
\begin{equation*}
    \left[A_{j2j1}\right]=\mathbf{U_j} \frac{\partial\mathbf{U_j}}{\partial x}+\mathbf{V_j}\frac{\partial \mathbf{U_j}}{\partial y}-\frac{\mathbf{U}-\mathbf{U_j}}{St_j}
\end{equation*}
\begin{equation*}
    \left[A_{j2j2}\right]=Q_{0,j}\left( i\alpha \mathbf{U_j} +\frac{\partial \mathbf{U_j}}{\partial x}+\mathbf{V_j} \textbf{\textit{D}}\right) +\left(\frac{Q_{0,j}}{St_j} +\underbrace{B_{j,j+1}Q_{0,j+1}}_{\text{vanishes when $j=N_s$}}\right) \textit{\textbf{I}}
\end{equation*}
\begin{equation*}
    \left[A_{j2j3}\right]=iQ_{0,j}\frac{\partial \mathbf{U_j}}{\partial y}
\end{equation*}
\begin{equation*}
    \left[A_{j2(j+1)1}\right]=-B_{j,j+1} \left(\mathbf{U_{j+1}}-\mathbf{U_j}\right) ~~~ \text{(vanishes when $j=N_s$)}
\end{equation*}
\begin{equation*}
    \left[A_{j2(j+1)2}\right]=-B_{j,j+1} Q_{0,j+1} \textit{\textbf{I}}~~~ \text{(vanishes when $j=N_s$)}
\end{equation*}
\begin{equation*}
    \left[A_{j3j1}\right]=\mathbf{U_j} \frac{\partial\mathbf{V_j}}{\partial x}+\mathbf{V_j}\frac{\partial \mathbf{V_j}}{\partial y}-\frac{\mathbf{V}-\mathbf{V_j}}{St_j}
\end{equation*}
\begin{equation*}
    \left[A_{j3j2}\right]=Q_{0,j}\frac{\partial \mathbf{V_j}}{\partial x}
\end{equation*}
\begin{equation*}
    \left[A_{j3j3}\right]=Q_{0,j}\left( -\alpha \mathbf{U_j} +i\mathbf{V_j} \textbf{\textit{D}}+i\frac{\partial \mathbf{V_j}}{\partial y}\right) +\left(i\frac{Q_{0,j}}{St_j} +\underbrace{iB_{j,j+1}Q_{0,j+1}}_{\text{vanishes when $j=N_s$}}\right) \textit{\textbf{I}}
\end{equation*}

\begin{equation*}
    \left[A_{j3(j+1)1}\right]=-B_{j,j+1} \left(\mathbf{V_{j+1}}-\mathbf{V_j}\right) ~~~ \text{(vanishes when $j=N_s$)}
\end{equation*}
\begin{equation*}
    \left[A_{j3(j+1)3}\right]=-iB_{j,j+1} Q_{0,j+1} \textit{\textbf{I}}~~~ \text{(vanishes when $j=N_s$)}
\end{equation*}
Note that the other matrices in the form of $[A_{axby}]$ that are not specifically listed above are zero-matrices.

Similarly, the matrix $\mathbf{B}$ can be constructed the same way, and its corresponding components are listed in detail below:
\begin{equation*}
    [B_{0201}]=i \textit{\textbf{I}} 
    \hspace{0.2in}
     [B_{0302}]=- \textit{\textbf{I}} 
\end{equation*}
for $j=1, 2, ..., N_s$:
\begin{equation*}
    [B_{j1j1}]=i \textit{\textbf{I}} 
    \hspace{0.2in}
     [B_{j2j2}]=iQ_{0,j} \textit{\textbf{I}} 
     \hspace{0.2in}
     [B_{j3j3}]=-Q_{0,j} \textit{\textbf{I}} 
\end{equation*}
And the remaining components are all zero-matrices.

\bibliographystyle{jfm}
\bibliography{yuval_ISF_2019,ref_zhi}

\end{document}